\documentclass[12pt,preprint]{aastex}
\slugcomment{Accepted to ApJ}

\def\blender{{\tt BLENDER}}
\def\kepler{\emph{Kepler}}
\def\kms{\ifmmode{\rm km\thinspace s^{-1}}\else km\thinspace s$^{-1}$\fi}
\def\ms{\ifmmode{\rm m\thinspace s^{-1}}\else m\thinspace s$^{-1}$\fi}
\newcommand{\logg}{\ensuremath{\log{g}}}
\newcommand{\rsun}{\ensuremath{R_\sun}}
\newcommand{\msun}{\ensuremath{M_\sun}}

\newcommand{\rstar}{\ensuremath{R_\star}}
\newcommand{\mstar}{\ensuremath{M_\star}}
\newcommand{\rjup}{\ensuremath{R_{\rm J}}}

\newcommand{\teff}{\ensuremath{T_{\rm eff}}}
\newcommand{\rhostar}{\ensuremath{\bar{\rho_\star}}}

\newcommand{\hd}{HD\,179070}

\shortauthors{Howell et al.}
\shorttitle{HD 179070's Super-Earth}

\begin{document}

\title{Kepler-21b: A 1.6R$_{\rm Earth}$ Planet Transiting the Bright Oscillating F Subgiant Star HD
179070
\footnote{Based in part on observations obtained at the W.~M.~Keck Observatory,
which is operated by the University of California and the California Institute of 
Technology, the Mayall telescope at Kitt Peak National Observatory, and the WIYN Observatory which is a joint facility
of NOAO, University of Wisconsin-Madison, Indiana University, and Yale University.}
}

\author{
Steve~B.~Howell\altaffilmark{1,2,36},
Jason~F.~Rowe\altaffilmark{2,34,36},
Stephen~T.~Bryson\altaffilmark{2}, 
Samuel~N.~Quinn\altaffilmark{3},
Geoffrey~W.~Marcy\altaffilmark{4},
Howard~Isaacson\altaffilmark{4},
David~R.~Ciardi \altaffilmark{5},
William~J.~Chaplin\altaffilmark{6},
Travis~S.~Metcalfe\altaffilmark{7},
Mario~J.~P.~F.~G.~Monteiro\altaffilmark{8},
Thierry~Appourchaux\altaffilmark{9},
Sarbani~Basu\altaffilmark{10},
Orlagh~L.~Creevey\altaffilmark{11,12},
Ronald~L.~Gilliland\altaffilmark{13},
Pierre-Olivier~Quirion\altaffilmark{14},
Denis~Stello\altaffilmark{15},
Hans~Kjeldsen\altaffilmark{16},
J\"orgen~Christensen-Dalsgaard\altaffilmark{16},
Yvonne~Elsworth\altaffilmark{6},
Rafael~A.~Garc\'ia\altaffilmark{17},
G\"unter~Houdek\altaffilmark{18},
Christoffer~Karoff\altaffilmark{6},
Joanna~Molenda-\.Zakowicz\altaffilmark{19},
Michael~J.~Thompson\altaffilmark{7},
Graham~A.~Verner\altaffilmark{20,6},
Guillermo~Torres\altaffilmark{3},
Francois~Fressin\altaffilmark{3},
Justin~R.~Crepp\altaffilmark{21},
Elisabeth~Adams\altaffilmark{3},
Andrea~Dupree\altaffilmark{3},
Dimitar~D.~Sasselov\altaffilmark{3},
Courtney~D.~Dressing\altaffilmark{3},
William~J.~Borucki \altaffilmark{2},
David~G.~Koch \altaffilmark{2},
Jack~J.~Lissauer\altaffilmark{2},
David~W.~Latham \altaffilmark{3},
Thomas~N.~Gautier III \altaffilmark{23},
Mark~Everett \altaffilmark{1},
Elliott~Horch \altaffilmark{25},
Natalie~M.~Batalha \altaffilmark{26},
Edward~W.~Dunham \altaffilmark{27},
Paula~Szkody\altaffilmark{28,36},
David~R.~Silva\altaffilmark{1,36},
Ken~Mighell\altaffilmark{1,36},
Jay~Holberg\altaffilmark{29,36},
Jer\^ome~Ballot\altaffilmark{30},
Timothy~R.~Bedding\altaffilmark{15},
Hans~Bruntt\altaffilmark{11},
Tiago~L.~Campante\altaffilmark{8,16},
Rasmus~Handberg\altaffilmark{16},
Saskia~Hekker\altaffilmark{6},
Daniel~Huber\altaffilmark{15},
Savita~Mathur\altaffilmark{7},
Benoit~Mosser\altaffilmark{31},
Clara~R\'egulo\altaffilmark{11,12},
Timothy~R.~White\altaffilmark{15},
Jessie~L.~Christiansen\altaffilmark{34}
Christopher~K.~Middour\altaffilmark{32}, 
Michael~R.~Haas\altaffilmark{2},
Jennifer~R.~Hall\altaffilmark{32},
Jon~M.~Jenkins \altaffilmark{34},
Sean~McCaulif\altaffilmark{32},
Michael~N.~Fanelli\altaffilmark{33}, 
Craig Kulesa\altaffilmark{35},
Don McCarthy\altaffilmark{35},
Christopher~E.~Henze\altaffilmark{2}
}

\altaffiltext{1}{National Optical Astronomy Observatory, 950 N. Cherry Ave., Tucson, AZ 85719}
\altaffiltext{2}{NASA Ames Research Center, Moffett Field, CA 94035}
\altaffiltext{3}{Harvard-Smithsonian Center for Astrophysics, 60 Garden Street, Cambridge, MA 02138}
\altaffiltext{4}{University of California, Berkeley, Berkeley, CA 94720}
\altaffiltext{5}{NASA Exoplanet Science Institute/Caltech Pasadena, 
CA 91125}
\altaffiltext{6}{School of Physics and Astronomy, University of Birmingham, Edgbaston, Birmingham, B15
2TT, UK}
\altaffiltext{7}{High Altitude Observatory and, Scientific Computing Division, National Center for
Atmospheric Research, Boulder, Colorado 80307, USA}
\altaffiltext{8}{Centro de Astrof\'\i sica, Universidade do Porto, Rua das Estrelas, 4150-762, Portugal}
\altaffiltext{9}{Institut d'Astrophysique Spatiale, Universit\'e Paris XI -- CNRS (UMR8617), Batiment
121, 91405 Orsay Cedex, France}
\altaffiltext{10}{Department of Astronomy, Yale University, P.O. Box 208101, New Haven, CT 06520-8101,
USA}
\altaffiltext{11}{Departamento de Astrof\'{\i}sica, Universidad de La Laguna, E-38206 La Laguna, Tenerife,
Spain}
\altaffiltext{12}{Instituto de Astrof\'{\i}sica de Canarias, E-38200 La Laguna, Tenerife, Spain}
\altaffiltext{13}{Space Telescope Science Institute, Baltimore, MD 21218}
\altaffiltext{14}{Canadian Space Agency, 6767 Boulevard de l'A\'eroport, Saint-Hubert, QC, J3Y 8Y9,
Canada}
\altaffiltext{15}{Sydney Institute for Astronomy (SIfA), School of Physics, University of Sydney, NSW
2006, Australia}
\altaffiltext{16}{Department of Physics and Astronomy, Aarhus University, DK-8000 Aarhus C, Denmark}
\altaffiltext{17}{Laboratoire AIM, CEA/DSM -- CNRS -- Universit\'e Paris Diderot -- IRFU/SAp, 91191
Gif-sur-Yvette Cedex, France}
\altaffiltext{18}{Institute of Astronomy, University of Vienna, A-1180, Vienna, Austria}
\altaffiltext{19}{Astronomical Institute, University of Wroc\l{}aw, ul. Kopernika, 11, 51-622 Wroc\l{}aw,
Poland}
\altaffiltext{20}{Astronomy Unit, Queen Mary, University of London, Mile End Road, London, E1 4NS, UK}
\altaffiltext{21}{California Institute of Technology Department of Astrophysics 1200 E. California Blvd.
Pasadena, CA, 91125}
\altaffiltext{22}{Massachusetts Institute of Technology, Cambridge, MA, 02139}
\altaffiltext{23}{Jet Propulsion Laboratory/California Institute of Technology, Pasadena, CA 91109}
\altaffiltext{24}{National Solar Observatory, 950 N. Cherry Ave, Tucson, AZ 85719}
\altaffiltext{25}{Southern Connecticut State University, New Haven, CT 06515}
\altaffiltext{26}{San Jose State University, San Jose, CA 95192}
\altaffiltext{27}{Lowell Observatory, Flagstaff, AZ 86001}
\altaffiltext{28}{Astronomy Department, University of Washington, Seattle, WA, 94065}
\altaffiltext{29}{LPL, University of Arizona, Tucson, AZ 85726}
\altaffiltext{30}{Laboratoire d'Astrophysique de Toulouse-Tarbes, Universit\'e de Toulouse, CNRS, 14 av
E. Belin, 31400 Toulouse, France}
\altaffiltext{31}{LESIA, CNRS, Universit\'e Pierre et Marie Curie, Universit\'e, Denis Diderot,
Observatoire de Paris, 92195 Meudon cedex, France}
\altaffiltext{32}{Orbital Sciences Corporation, NASA Ames Research Center, Moffett Field,
CA 94035}  
\altaffiltext{33}{Bay Area Environmental Research Inst., NASA Ames Research Center, Moffett
Field, CA 94035}  
\altaffiltext{34}{SETI Institute, Mountain View, CA 94043}
\altaffiltext{35}{Steward Observatory,University of Arizona, Tucson, AZ 85726}
\altaffiltext{36}{Visiting Astronomer, Kitt Peak National Observatory, National Optical
Astronomy Observatory, which is operated by the Association of Universities for
Research in Astronomy (AURA) under cooperative agreement with the National Science
Foundation.}


\keywords{planetary systems --- stars: individual (HD 179070: KIC3632418) --- 
stars: oscillations --- stars: interiors --- stars:
late-type -- stars: activity -- stars: magnetic field
--- techniques: photometric --- Facilities: The Kepler Mission}

\begin{abstract} 

We present {\it Kepler} observations of the bright (V=8.3), oscillating star HD 179070. 
The observations show transit-like events which
reveal that the star is orbited every 2.8 days by a small, 1.6 $R_{\rm Earth}$ object.
Seismic studies of HD 179070 using short cadence {\it Kepler} observations 
show that HD 179070 has a frequency-power spectrum consistent with solar-like oscillations
that are acoustic p-modes. Asteroseismic analysis provides robust values for the mass and radius
of HD 179070, 1.34$\pm$0.06 M$_{\odot}$  and 1.86$\pm$0.04 R$_{\odot}$ respectively, as well as 
yielding an age of 2.84$\pm$0.34 Gyr for this F5 subgiant.
Together with ground-based follow-up observations, analysis of the {\it Kepler} light curves and 
image data, and blend scenario models, we conservatively show at the $>$99.7\% confidence 
level (3$\sigma$) that the transit event is caused by a 
1.64$\pm$0.04 R$_{\rm Earth}$ exoplanet in a 2.785755$\pm$0.000032 day orbit. 
The exoplanet is only 0.04 AU away from the star and our spectroscopic observations provide 
an upper limit to its mass of $\sim$10 $M_{\rm Earth}$ (2-$\sigma$). 
HD 179070 is the brightest exoplanet host star yet discovered by {\it Kepler}.

\end{abstract}

\section{Introduction}

The NASA {\it Kepler} mission, described in  Borucki et al. (2010a), surveys a large area of the
sky spanning the boundary of the constellations Cygnus and Lyra. The prime mission goal is to 
detect transits by exoplanets as their orbits allow them to pass in front of their
parent star as viewed from the Earth. 
Exoplanets, both large and small and orbiting bright and faint stars, 
have already been detected by the {\it Kepler} mission
(see Borucki et al. 2010b; Koch et al. 2010a; Batalha et al. 2010; Latham et al. 2010; 
Dunham et al. 2010; Jenkins et al. 2010a; Holman et al. 2010; 
Torres et al. 2010; Howell et al. 2010; Batalha et al. 2010; Lissauer et al. 2011)
and many more exoplanet candidates are known (Borucki et al. 2011).

We report herein on one of the brightest targets in the Kepler field, HD 179070 (KIC 3632418, KOI 975).
This V=8.3 F6IV star has been observed prior to the Kepler mission as a matter of course for
essentially all of the Henry Draper catalogue stars. Catalogue
information\footnote{http://simbad.u-strasbg.fr/simbad/sim-fid} for HD179070 provides a T$_{\rm eff}$ of
6137K (spectral type F6 IV), an Hipparcos distance of 108$\pm$10 pc, [Fe/H]=-0.15, an 
age of 2.8 GYr, a radial velocity of -28 km/sec and E(b-y)=0.011 mag.
The age, metallicity, and color excess reported in the catalogue come from the photometric study by 
Nordstr\"{o}m,  et al. (2004). 
nothing special to distinguish it from any other stars in the HD catalogue.

HD 179070 is strongly saturated in the {\it Kepler} observations.
Techniques to obtain good photometry from saturated stars
are employed by the {\it Kepler} project and have been used to analyze light
curves of other {\it Kepler} saturated stars (e.g., Welsh et al. 2011)
While a star will saturate
the CCD detectors on {\it Kepler} if brighter than V$\sim$R$\sim$11.5, collecting all the pixels which
contain the starlight, including the bleed trail, allow a photometric study to be performed
(see Gilliland et al. 2010a).
{\it Kepler} light curves of HD179070 from the early data as well as subsequent quarters of
observation have shown a repeatable transit-like event in the light curve. We will examine the
nature of this periodic signal using Quarter 0 to Quarter 5 {\it Kepler} observations 
and will show that it is caused by a small exoplanet
transiting HD 179070 with a period near 2.8 days. We will refer to the host star as HD179070 and
the exoplanet as Kepler-21b throughout this paper.

Our analysis for this exoplanet follows the very complete, tortuous validation path fully 
described in Batalha et al.
(2010). The interested reader is referred to that paper for the detailed procedures
which we do not completely repeat herein.
Section 2 will discuss the {\it Kepler} observations including both the light curves and the pixel image
data. \S3 \& \S4 discuss ground-based speckle and 
spectroscopic follow-up observations we have obtained and 
Section 5 presents the asteroseismic results.
Using a full analysis of the observations, we produce a transit model fit for 
the exoplanet, determine many observational properties for it, and end with a discussion 
of our results.

\section{{\it Kepler} Observations}

\subsection{{\it Kepler} Photometry}

The {\it Kepler} mission and its photometric performance since launch are
described in  Borucki et al. (2010a) while the CCD imager on-board {\it
Kepler} is described in  Koch et al. (2010b) and van Cleve (2008). 
The {\it Kepler} observations of HD 179070 used herein consist of data covering a time period of 
460 days or {\it Kepler} observation Quarters 0 through 5 (JD 2454955 to 2455365).
The photometric observations
were reduced using the {\it Kepler} mission data pipeline (Jenkins et al.
2010b) and then passed through various consistency checks and exoplanet
transit detection software as described in Van Cleve (2009) and Batalha et al. (2011). 
Details of the Kepler light curves and the transit model fitting procedures
can be found in Howell et al. (2010), Batalha et al. (2011) and Rowe et al. (2006).

The top panel of Fig. 1 shows
the raw {\it Kepler} light curve of HD 179070 where the larger, low frequency modulations
do not likely represent real changes in the star.
Thermal jumps in the focal plane temperature near day 100, 120, and 370 are apparent 
(see Van Cleve 2009) due to safeing events in Q2 and Q5.
Normalized and phase-folded light curves are produced 
and the transit event in the phased light curve is
modeled in an effort to understand the transiting object.
The middle panel of Fig. 1 shows a blow up (as highlighted in yellow) of a section of the raw light curve
and represents a typical normalized result.
The dotted lines mark the individual transit events and the entire {\it Kepler} 
light curve shown here contains 164 individual transits.
Taking random quarters of the total light curve and binning them on the transit period
provide consistent results as shown in the bottom panel of Fig. 1 albeit of lower S/N.
The bottom panel (Fig. 1) shows the entire phase folded light curve after 
detrending and binning all
available data.  Each bin has a width of 30 minutes and the red curve shows
our transit model fit to the data. 
Points marked with 'o's show where the exoplanet occultation would
occur for a circular orbit (i.e., light curve phased at 0.5)
or where evidence of a secondary eclipse would be
seen if the event arises from a blended, false-positive
eclipsing binary.

Due to the sparse (30-minute) sampling of the image data over the 3.4 hour long 
transit as well as the 
low S/N of any individual transit event, a search for transit timing variations 
produced no detectable periodic signal with an amplitude of 8 mins 
or greater and no significant deviations at all for any period.

\subsection{{\it Kepler} Images and False Positive Analysis}

HD 179070 is saturated in {\it Kepler} images. In a typical quarter the 
photometric aperture covers 128 pixels, of
which 39 are typically saturated in each image.
A direct pixel image of HD 179070 in quarter 5 is shown in Figure~\ref{fig:directImage}.

False positive identification for unsaturated {\it Kepler} targets proceeds by forming an average
difference
image per quarter by subtracting an average of the in-transit pixel values from average nearby
out-of-transit pixel values.  The resulting difference image is used to compute a
high-accuracy centroid of that difference image (Torres et al. 2010).
For unsaturated targets the difference image provides a star-like image at the actual location of the
transiting object.  The high-accuracy centroid is computed by performing a fit of the difference image to
the {\it Kepler}
pixel response function (PRF; see Bryson et al. 2010).  
When the difference images are well behaved, the centroid
can have precisions on the order the PSF scale divided by
the photometric signal to noise ratio of the transit signal.
For Kepler-21b the SNR per quarter of about 25 would support
centroiding precisions for sources near the primary of
about 0.06 pixels.  Having lost spatial resolution due to
saturation the centroiding capability will be degraded
compared to this.

The high degree of saturation of HD 179070 prevents the direct application of the above centroiding
technique, but we believe that a PRF fitted centroid to the non-saturated pixels in 
HD 179070's wings provide useful, albeit less accurate, results. 

Figure~\ref{fig:diffImage} shows the average difference image for Quarter 5, which 
is typical of other quarters.
The pixel values which change during a transit are at the ends of the saturated columns 
as the amount of saturation
that spills out of these ends decreases during transit.  We also detect 
the (in phase) transit signal in the stellar wings around
the
core to the right of the saturated columns. 
Some saturated pixels become brighter on average during the transits which we ascribe 
to negative difference values in the pixel-level systematics linked to 
the non-linear behavior of saturated pixels and to large outliers caused by image motion
events.
Such negative difference image pixel values are commonly associated with 
saturated {\it Kepler} targets in which we observe shallow transits.

We can enhance our view of the stellar wings around the saturated core of HD 179070
by setting pixels in columns that have saturated pixels in the
direct image to zero in the
left panel of Figure~\ref{fig:noSatDiffImage}.  The right panel shows the modeled 
difference pixel image
created by simulating the transit on HD 179070 using the PRF model of Bryson et al. (2010)
and the measured transit depth, similarly
setting the saturated columns to zero.  We see that for the stellar wings on the right side
there is a reasonable qualitative match between the modeled
difference image and the observed pixel values. However, on the left side, 
the observed wings have slightly negative values. Since we can not know the exact center
of the stellar image due to the many saturated pixels, 
we expect that the PRF-fitted
centroid will have some x,y bias. These negative values appear for all quarters, 
but their locations
around the core vary significantly from quarter to quarter, indicating that the
PRF-fitting bias varies from quarter to quarter.

The PRF fit is performed via Levenberg-Marquardt minimization of the $\chi^2$ difference between the
modeled and observed difference values, using non-saturated pixels in the model image whose value exceeds
$10^{-4}$ of the summed model pixel values.
This fit is done on both the difference and direct image, allowing us to
compare the centroid of the difference image with the measured centroid of HD 179070 using the same
pixels.
The resulting in-transit pixel offsets are given in Table~\ref{tab:centroids}, with their formal
uncertainties, for quarters
1, 3, 4 and 5 (no quarter 2 data are available).  These
uncertainties do not include the expected but unknown fit bias due to the negative values in the observed
difference image.
In quarter 1 we see centroid offsets exceeding a pixel, but this is almost certainly 
due to significant image-motion related to
pixel-level systematics that were eliminated in later quarters (Jenkins et al. 2010c).
In the remaining quarters the offsets are smaller,
particularly in quarters 4 and 5 where the offsets are less than a pixel, 
and there is no consistency, as expected, in the offsets.

To gain confidence in PRF fitting using pixels in the stellar wings and in using the {\it Kepler}
image data for HD 179070 as a way to set limits on possible background sources which could make this a false
positive, a series of modeled centroids was produced.
The source of the false  positive event
was modeled to be caused by a very dim variable test star (e.g., a background eclipsing binary) 
with a {\it Kepler} magnitude of 18 in which a
variation (eclipse) was added with a depth of 50\%. 
Models were produced with the faint test star placed at various pixel positions near the expected center of
HD 179070 but such that some of the test stars' light would spill into the unsaturated wings. 
While the models did not exactly reproduce the observed data due to unknown systematics, the location of the 
saturated columns was faithfully reproduced when the transit was assumed to be on HD 179070.
When the transit was placed on a background star offset by more than half a pixel in column the 
saturated column clearly became inconsistent with the data.  When the transiting object is offset by a pixel
or more in the row direction the location of the wings in the model data became similarly 
inconsistent with the observed data. In this way the models show
that the observed difference image can clearly rule out the possibility that the transiting object 
is located more than a pixel (4 arcsec) from HD 179070.  
Row offsets displace the model stellar wings in a difference image
by large fractions of a pixel relative to the observed difference images.
The effect of a column offset is shown in Figure~\ref{fig:colOffsetDiffImage}, where we see that a column
offset
of one pixel (or more) causes the transit signal in the difference image to disappear from one or more of 
the saturated columns again not consistent with the {\it Kepler} observations.
Based on our PRF model results, we are confident that the source of the 
transiting object is $\le$1 pixel of the x,y position of HD 179070. 


\section{High Resolution Imaging}

\subsection{Speckle Observations}

A major part of the {\it Kepler} follow-up program (Batalha et al. 2010) used to 
find false positives as well as provide ``third light" information to aid in {\it Kepler} image
analysis is speckle imaging.
We perform our speckle observations at the 3.5-m WIYN telescope located on Kitt Peak
where we make use of the Differential
Speckle Survey Instrument, a recently upgraded speckle camera described in
Horch et al. (2010).
Our speckle camera provides simultaneous observations in two filters
by employing a dichroic beam splitter and two identical EMCCDs as the imagers. We generally observe 
simultaneously in
"V" and "R" bandpasses where "V"  has a central wavelength of 5620\AA~,
and "R" has a central wavelength of 6920\AA~, and each filter has a FWHM=400\AA.
The details of how we obtain, reduce, and analyze the speckle results and specifics about how 
they are used eliminate false
positives and aid in transit detection are described in Torres et al. (2010), Horch et al.
(2010), and Howell et al. (2011).

The speckle observations of HD 179070 were obtained on 17 Sept. 2010 UT 
and consisted of one  set of 1000, 40 msec speckle images. Our
R-band reconstructed image is shown in Figure 6
and along with a nearly identical V-band reconstructed image
reveals no companion star near HD 179070 within the annulus from 
0.05 to 1.8 arcsec to a limit of (5$\sigma$) 5.3 magnitudes fainter than the target star
(that is brighter than R$\sim$13.6). 
At the distance of HD 179070 (d$\sim$108 pc), this annulus corresponds to distances of 
5.4 to 194 AU from the star. We note that
any stellar companions or massive exoplanets ($\sim$20 Earth masses or more) 
inside 5.4 AU would be easily detectable in the radial velocity signature
(see \S4). 
We note that while reaching to 5 magnitudes fainter than the target star eliminates bright 
companions and some fraction of low-mass faint associated
companions, it does not completely rule out the probable larger population of faint background EBs (see \S7).

\subsection{MMT AO Imaging}

AO images of HD 179070 were obtained on 2010 September 23 UT using the
ARIES instrument on the 6.5-m MMT. ARIES is a near-infrared
diffraction-limited imager, and was operated in the F/30 mode (0.02"
per pixel) in both the $J$ and $Ks$ filters. The combined Ks image was
created by combining 18 images: one 1-s and 17 0.9-s exposures, with
two initial pointings at each exposure time and a raster of 4 4-point
dithers with a jittered 2" offset between each position. The combined
J image was similarly made from 17 0.9-s exposures (one single and
sixteen dithered images). The images were shifted, sky-subtracted, and
combined using xmosaic in the IRAF package xdimsum.
The final AO images are shown in Figure 7.

The seeing was relatively poor, with image FWHM of 0.33" in $J$ and
0.29" in $Ks$. A faint companion is detectable in the $Ks$
data, and hinted at in the $J$ data. PSF fitting, using an analytical
Gaussian model, found a magnitude difference of $\Delta-Ks = 3.7 +/-
0.1$, while a lower limit was estimated for $\Delta-J > 4.0$. The
separation between the two components was found to be 0.7" +/- 0.05",
with an approximate position angle of 135 degrees east of north.

The faint companion was detected right at the limit of the ARIES
frame, and at a distant of just over 2 FWHM. At the object's distance,
0.7", the estimated $Ks$ detection limit for additional companions was
4.2 mag fainter (3.6 mag in $J$), increasing to 5.7 mag in $Ks$ (5.1
in $J$) at 1", 7.5 mag in $Ks$ (7.2 in $J$) at 2", and 8 mag in $Ks$
(7.8 in $J$) beyond 4". In order to get better magnitude limits and a
firm $J$-band detection, additional AO images were acquired using
Keck.

\subsection{Keck AO Imaging}

Near-infrared adaptive optics imaging of HD 179070 was obtained on the night
of 22 February 2011 and 23 February 2011 with the Keck-II  telescope and
the NIRC2 near-infrared camera behind the natural guide star adaptive
optics system.  NIRC2, a $1024\times1024$ HgCdTe infrared array, was
utilized in 9.9 mas/pixel mode yielding a field of view of
$\approx10\arcsec$.  Observations were performed on first night in the 
K-prime filter ($K^\prime$; $\lambda_0 = 2.124 \mu$m; $\delta\lambda =
0.351 \mu$m), and on the second night in the $J$ filter ($\lambda_0 = 1.248
\mu$m; $\delta\lambda = 0.163\mu$m).  A single $K^\prime$ data frame was
taken with an integration time of 2 seconds and 10 coadds; 10 frames were
acquired for a total integration time of 200 seconds.  A single $J$ data
frame was taken with an integration of 0.18s and 20 coadds; 10 frames were
acquired  for a total integration time of 36 seconds.  The weather on the
night of the observations was poor with occasional heavy clouds.

The individual frames were background-subtracted and flat-fielded into a single
final image for each filter. The central core of the resulting point spread
functions had a width of $FWHM = 0.07\arcsec$ ($\approx 7$ pixels) at J and
$FWHM = 0.09\arcsec$ ($\approx 9$ pixels) at $K^\prime$. The final coadded images are
shown in Figure~\ref{fig:keckAO}.  A faint source is detected
$0.75\arcsec$ from the primary target at a  position angle of $PA =
129^\circ$ east of north. The source is fainter than the primary target by
$\Delta J = 4.70 \pm 0.05$ mag and $\Delta K^\prime = 3.95 \pm 0.05$ mag. 
No other sources were detected within $5\arcsec$ of the primary target.  

If the faint companion is a dwarf star, the $J-Ks$ color implies that it is
a very late M-dwarf ($\sim$M5-M8; see Leggett et al. 2002; Ciardi et al. 2011) 
and would be at a distance of approximately 15$\pm$8 pc.  
If the companion is a giant star, the
$J-Ks$ color implies that it is an M0 giant and would have an approximate
distance of 10 kpc.  
Appendix A discusses in detail our use of the near-IR AO observation to convert
the companion's brightness into K$_p$ = 14.4$\pm$0.2.
The maximum line of sight extinction to the faint
companion (as determined from the IRAS/DIRBE dust maps;
see Schlegel et al., 1998) 
is $A_V \approx 0.5$ mag, which corresponds to an $E(J-Ks)
\approx 0.09$ mag; such an excess would only change the implied spectral
type by a single subclass (M5 dwarf and K5 giant) and would not appreciably
change the derived distances.  
The red dwarf can be made much earlier in spectral type (and thus located at the same distance of HD 179070)
but this would require significant reddening along the exact line of sight to the star 
which is ruled out based on the color excess listed for HD 179070 in the literature (E(b-y)=0.011). 
The primary target has a Hipparcos distance
of $108\pm 10$ pc; thus, the faint companion, whether it is a dwarf or
giant, is not physically associated with the primary target.

Source detection completeness was estimated by randomly inserted fake
sources of various magnitudes in steps of 0.5 mag and at varying distances
in steps of 1.0 FWHM from the primary target. Identification of sources was
performed both automatically with the IDL version of DAOPhot and by eye.
Magnitude detection limits were set when a source was not detected by the
automated FIND routine or was not detected by eye.  Within a distance of
$1-2$ FWHM, the automated finding routine often failed even though the eye
could discern two sources, particularly since the observations were taken
in poor weather conditions.  A summary of the detection efficiency as a
function of distance from the primary star is given
Table~\ref{tab:keckAO}.  Beyond $\approx 0.7\arcsec$, the detection
limit is $\approx 6$ magnitudes fainter than the target.

\section{Spectroscopic Observations}

Optical spectroscopy for HD 179070 was obtained by three ground-based telescopes as part of the
{\it Kepler} mission follow-up program (Batalha et al. 2010). These observations include early
reconnaissance spectra to assess stellar identification, rotation, and to make a first check on
binary as well as detailed high precision radial velocity work using Keck HIRES.
We compare all of our ground-based spectroscopic determined values in Table 2.

Reconnaissance spectra of HD179070 were obtained on 13 \& 16 Sept. 2010 UT 
at the Kitt Peak 4-m telescope using the RCSpec instrument. 
The 4-m RCSpec setup used a 632 l/mm grating (KPC-22b in second
order) with a 1 arcsec slit to provide a mean spectral resolution of
1.6\AA~per resolution element across the full wavelength range of 3750-5100\AA. The spectra
were reduced in the normal manner with observations of calibration lamps
and spectrophotometric stars (obtained before and after each sequence) and
bias and flat frames collected each afternoon. 
Each fully reduced 4-m spectrum (see Figure 8) was cross-correlated and $\chi^2$ fit 
to both the entire MK standard stars digitally
available in the ``Jacoby Atlas" (Jacoby et al. 1984; covers all spectral and luminosity types) 
as well as to a suite of stellar models (ranging in T$_{\rm eff}$ from 3500K to 7000K, log g from
1.0 to 5.0, and solar metallicity)
available through the Spanish Virtual Observatory\footnote{http://svo.laeff.inta.es/}.

Spectral type, luminosity class, and other stellar parameters were provided by the best fit
match and both of the 4-m spectra gave consistent results: F4-6 IV star with 
T$_{\rm eff}$ = 6250$\pm$250K, log g = 4.0$\pm$0.25, and metal poor ([Fe/H] = -0.15). 
No relevant $vsini$ information was available from the moderate resolution (R$\sim$5000) 4-m spectra.

As is common procedure for the Kepler Mission, all exoplanet candidate stars also receive
high resolution, low signal to noise (S/N), spectroscopic observations to
identify easily recognizable astrophysical false positives. One or two
correctly timed spectra can help rule out many types of false
positives, including single- and double-lined binaries, certain types
of hierarchical triples and even some background eclipsing binaries,
all of which show velocity variations and/or composite spectra that
are readily detectable by the modest facilities used for these
reconnaissance observations. We also use these spectra to estimate the
effective temperature, surface gravity, metallicity, and rotational
and radial velocities of the host star. Below is a brief description
of the instrument, the data reduction, and the analysis performed in this step.

We used the Tillinghast Reflector Echelle Spectrograph (TRES; Furesz
2008) on the 1.5m Tillinghast Reflector at the Fred L. Whipple
Observatory on Mt. Hopkins, AZ to obtain a high resolution, low S/N
spectrum of HD 179070 (S/N$\sim$7 per resolution element, 360 second exposure) on 28 Sep 2010
UT. The observation was taken with the medium fiber on TRES, which has
a resolving power of $\lambda$/d$\lambda$ $\sim$ 44,000 and a wavelength coverage
of 3900-8900 angstroms. The spectrum was extracted and analyzed
according to the procedures outlined by Buchhave et
al. (2010). Cross-correlations were performed against the grid of CfA
synthetic spectra, which are based on Kurucz models calculated by John
Laird and rely on a line-list compiled by Jon Morse. The template with
the highest correlation coefficient yields an estimate of the stellar
parameters: T$_{\rm eff}$ = 6250$\pm$125 K, log(g) = 4.0$\pm$0.25, and 
Vrot = 8$\pm$ 1 km/s. The errors correspond to half of the grid
spacing, although they neglect possible systematics, e.g., those
introduced in the event the metallicity differs from solar. We find
the absolute radial velocity to be V$_{rad}$ = -19.1$\pm$0.3 km/s.

Finally, HD 179070 was subjected to RV measurements with Keck (Vogt et al. 1994, Marcy et al. 2008).
We note that the Ca-II K-line shows virtually no chromospheric reversal, giving an $S$ value of
0.14, placing HD 179070 among the quietest G stars.
We have obtained 14 RV measurements with the Keck telescope and HIRES spectrometer (R=60,000). 
The exposure time for each spectrum was established by use of an exposure meter such that all would 
yield a consistent S/N
and thus very similar RV precision. The Keck HIRES exposures were 150$\pm$30 sec in all cases. 
Each observation consisted of a triplet of
exposures with each exposure having a signal-to-noise of 210 per pixel and an internal RV error of $\sim$2 m s$^{-1}$. 
We determined the RV from each exposure and took the weighted mean as the final RV for each triplet.  
The internal error (a typical uncertainty for each exposure) was based on the iodine line fits for 
each exposure.

We adopted a jitter of 5 m
s$^{-1}$, typical for stars of such spectral type and rotational Vsini, adding the jitter in quadrature to the internal errors
that were $\sim$2 m s$^{-1}$.   
Low-gravity F6 stars are well known to exhibit jitter of $\sim$5-10 m/sec (Isaacson \& Fischer 2010)
due presumably to photospheric velocity fields 
but the exact origin remains unclear. F-type main sequence stars engage in a
quasi-stable $\delta$ Scuti-like phenomenon, a feature 
not seen in the G and K stars. 
Table 3 gives the times and the velocity measurements and their uncertainties.
The jitter was not added to these uncertainties in quadrature, offering the reader a
chance to see the uncertainties pre-jitter.
The resulting 14 RVs had a standard deviation of 5.6 m s$^{-1}$, consistent with the expected total
errors.  

In Figure 9, we show the velocities of HD 179090 measured with the HIRES spectrometer on the Keck 1
telescope during $\sim$90 days in 2010 and 2011.  See Jenkins et al. (2011) and Batalha et al.
(2011) for a detailed explanation of the (standard) method we used with the iodine cell to make these
Doppler measurements. 
The velocities in Figure 9 present no clear long term variability on time scales of weeks or months.   A
periodogram of the velocities reveals no significant power at any period from 0.5 d to the duration of the RV
observations, 85 dy, including periodicity in the velocities at the transit period of
2.7857 d.
The Keck RVs are shown in Figure 10 as a function of known orbital phase (from the Kepler photometric light curve).
The RV variation measured shows no modulation coherent with the orbital phase and is consistent with no change at all 
within the uncertainties.

We performed a standard LTE spectroscopic analysis (Valenti and Piskunov 1996; Valenti and Fischer 2005)
of a high resolution template spectrum from Keck-HIRES 
to derive an effective temperature, $\teff$  = 6131 K,
surface gravity, $\logg$ = 3.9 (cgs), metallicity, [Fe/H]=-0.05 and 
$Vsini$ = 7.5 km/sec. Ground-based high spectral resolution
support observations of HD 179070 were also performed by the 
Kepler Asteroseismic Science Consortium (KASC).
Two teams obtained estimates for [Fe/H]; Molenda-Zakowicz et al. (2010) listed two values (-0.15 and -0.23, 
and [Fe/H] = -0.15 was determined by H. Bruntt (priv. comm.) using NARVAL at Pic du Midi 
and the VWA analysis package. 
We adopt [Fe/H]=-0.15 for HD 179070 in this paper using the more reliable value recently obtained by Bruntt.

\section{Asteroseismic analysis}
\label{sec:astero}

\subsection{Estimation of asteroseismic parameters}
\label{sec:extract}

HD 179070 was observed for one month by \emph{Kepler} at a short cadence
of 58.85\,s. A time series was prepared for asteroseismic analysis in
the manner described by Garc\'ia et al. (2011). Fig.~\ref{fig:powspec}
plots the frequency-power spectrum of the prepared time series, which
shows a beautiful pattern of peaks due to solar-like oscillations that
are acoustic (pressure, or p) modes of high radial order, $n$. 
The
observed power in the oscillations is modulated in frequency by an
envelope that has an approximately Gaussian shape. The frequency of
maximum oscillation power, $\nu_{\rm max}$, has been shown to scale to
good approximation as $gT_{\rm eff}^{-1/2}$ (Brown et al. 1991;
Kjeldsen \& Bedding 1995), where $g$ is the surface gravity and
$T_{\rm eff}$ is the effective temperature of the star.  
The $l$ identification, from visual inspection of the mode 
pattern in the frequency-power spectrum, is unambiguous and the $n$ 
identification followed from the best-fitting to stellar evolutionary 
models (see below). The most
obvious spacings in the spectrum are the large frequency separations,
$\Delta\nu$, between consecutive overtones $n$ of the same spherical
angular degree, $l$. These large separations scale to very good
approximation as $\left< \rho \right>^{1/2}$, $\left< \rho \right>
\propto M/R^3$ being the mean density of the star, with mass $M$ and
surface radius $R$ (e.g. see Christensen-Dalsgaard 1993).

Here, seven teams estimated the average large separation, $\left<
\Delta\nu \right>$, and $\nu_{\rm max}$, using automated analysis
tools that have been developed, and extensively tested (e.g., see
Campante et al. 2010a; Christensen-Dalsgaard et al. 2010; Hekker et
al. 2010; Huber et al. 2009; Karoff et al. 2010; Mosser \& Appourchaux
2009; Mathur et al. 2010a; Roxburgh 2009) for application to the large
ensemble of solar-like oscillators observed by \emph{Kepler} (Chaplin
et al. 2010,2011; Verner et al. 2011). A final value of each parameter was
selected by taking the individual estimate that lay closest to the
average over all teams. The uncertainty on the final value was given
by adding (in quadrature) the uncertainty on the chosen estimate and
the standard deviation over all teams. We add that there was excellent
consistency between results, and no outlier rejection was
required. The final values for $\left< \Delta\nu \right>$ and
$\nu_{\rm max}$ were $60.86 \pm 0.55\,\rm \mu Hz$ and $1153 \pm
32\,\rm \mu Hz$, respectively. We did not use the average frequency
separation between the $l=0$ and $l=2$ modes (often called the small
separation) in the subsequent modeling because HD 179070 turns out to
be a subgiant (as indicated by, for example, the size of $\left<
\Delta\nu \right>$), and in this phase of evolution the parameter
provides little in the way of additional constraints given the modest
precision achievable in it from one month of data (see Metcalfe et
al. 2010; White et al., 2011).

Use of individual frequencies increases the information content
provided by the seismic data for making inference on the stellar
properties.  Six teams provided estimates of individual frequencies,
applying ``peak bagging'' techniques developed for application to
CoRoT (Appourchaux et al. 2008) and \emph{Kepler} data (e.g., see
Metcalfe et al. 2010; Campante et al. 2010b; Mathur et al. 2010b;
Fletcher et al. 2010). We implemented the procedure outlined in
Campante et al. (2011) and Mathur et al. (2011) to select from the
six sets of estimated frequencies one set that would be used to model
the star. This so-called ``minimal frequency set'' contains estimates
on modes for which a majority of the teams' estimates were retained
after applying Peirce's criterion (Peirce 1852; Gould 1855) for
outlier rejection. Use of one of the individual sets, as opposed to
some average over all sets, meant that the modeling could rely on an
easily reproducible set of input frequencies. The selected frequencies
are listed in Table~\ref{tab:freqs}.

\subsection{Estimation of stellar properties}
\label{sec:prop}

We adopted two approaches to estimate the fundamental properties of
HD 179070.  In the first we used a grid-based approach, in which
properties were determined by searching among a grid of stellar
evolutionary models to get a best fit for the input parameters, which
were $\left< \Delta\nu \right>$, $\nu_{\rm max}$, and $T_{\rm eff} =
6131 \pm 44\,\rm K$, from spectroscopic observations made on the Keck
telescope in support of the HD 179070 analysis, and [Fe/H]$=-0.15 \pm
0.06$ from Molenda-\.Zakowicz et al. (2011). Descriptions of the
grid-based pipelines used in the analysis may be found in Stello et
al. (2009), Basu et al. (2010), Quirion et al. (2010) and Gai et
al. (2011).

In the second approach, the individual frequencies $\nu_{nl}$ were
analyzed by the Asteroseismic Modeling Portal (AMP), a web-based tool
tied to TeraGrid computing resources that uses the Aarhus stellar
evolution code ASTEC (Christensen-Dalsgaard 2008a) and adiabatic
pulsation code ADIPLS (Christensen-Dalsgaard 2008b) in conjunction
with a parallel genetic algorithm (Metcalfe \& Charbonneau 2003) to
optimize the match to observational data (see Metcalfe et al. 2009,
Woitaszek et al. 2009 for more details).

Each model evaluation involves the computation of a stellar evolution
track from the zero-age main sequence (ZAMS) through a mass-dependent
number of internal time steps, terminating prior to the beginning of
the red giant stage. Exploiting the fact that $\left< \Delta\nu
\right>$ is a monotonically decreasing function of age (see Metcalfe
et al. 2009, and references therein), we optimize the asteroseismic
age along each evolution track using a binary decision tree. The
frequencies of the resulting model are then corrected for surface
effects following the prescription of Kjeldsen et al. (2008). A
separate value of $\chi^2$ is calculated for the asteroseismic and
spectroscopic constraints, and these values are averaged for the final
quality metric to provide more equal weight to the two types of
observables.  The optimal model is then subjected to a local analysis
employing a modified Levenberg-Marquardt algorithm that uses singular
value decomposition (SVD) to quantify the values, uncertainties and
correlations of the final model parameters (see Creevey et al. 2007).

\subsection{Results on stellar properties}
\label{sec:res}

Both approaches to estimation of the stellar properties yielded
consistent results on the mass and radius of the star.  The final
estimates are $M=1.34 \pm 0.01$(stat)$\pm 0.06$(sys)$\,\rm M_{\odot}$
and $R=1.86 \pm 0.02$(stat)$\pm 0.04$(sys)$\,\rm R_{\odot}$.  The
statistical uncertainties come from the SVD analysis of the
best-fitting solution to the individual frequencies. The spreads in
the grid-pipeline results -- which reflect differences in, for
example, the evolutionary models and input physics -- were used to
estimate the systematic uncertainties.

The grid pipelines, which used only the average seismic parameters,
showed (in some cases) two possible solutions for the age of the star
(one around 3\,Gyr and another around 4\,Gyr or higher). Use of the
individual frequencies resolved this ambiguity by giving a
best-fitting solution that clearly favored the younger model, the
best estimate of the age being $\tau = 2.84 \pm 0.10$(stat)$\pm
0.33$(sys)$\,\rm Gyr$. Fig.~\ref{fig:ech} shows good agreement between
the frequencies of the best-fitting model and the observed
frequencies. This \'echelle diagram (e.g., see Grec et al. 1983) plots
the frequencies against those frequencies modulo the average large
frequency separation of $60.86\,\rm \mu Hz$. Overtones of the same
spherical degree $l$ are seen to align in near-vertical ridges. The
observed frequencies from Table~\ref{tab:freqs} are plotted in black,
along with their associated $1\sigma$ uncertainties; while the
best-fitting model frequencies are plotted in color.

\section{{\it Kepler} Photometry Transit Fits}\label{tfits}

The raw Q1-Q5 light curve is presented in the top panel of Figure 1.
The trends observed on various timescales are a combinination of
astrophysical phenomenon and instrumental artifacts.  These trends
were removed from the lightcurve as our lightcurve model did not
account for such effects.  The photometric time series was prepared
for modeling by independently detrending each quarterly {\it Kepler}
time-series.  A cubic polynomial was fitted and removed and then
filtered with a 5-day running median.    Any observations that
occurred during transit where masked out during the calculation of the
polynomial fits or medians.  Our model fits for the physical and
orbital parameters of the planetary system.  The transit shape was
described by the analytic  formulae of Mandel \& Agol 2002.  We
adopted a non-linear limb-darkening law (Claret).  Coefficients were
calculated by convolving Atlas-9 spectral models with the Kepler
bandpass using the adopted estimate of \teff\  from Table 2 and the
asteroseismic value of \logg.  Limb darkening coefficients are held
fixed for all transit fits. We assumed a Keplerian orbit for the
planet with zero eccentricity.  Our model fitted for the period (P),
epoch (T0), impact parameter (b), the mean stellar density (\rhostar),
the ratio of the planet and star radii (R$_p$/\rstar), radial velocity
amplitude (K) and the radial velocity zero point ($\gamma$).  A set of
best fit model parameters was constructed by fixing \rhostar\ to the
best value from asteroseimology.  Minimization of Chi-squared was
found using a Levenberg-Marquart method allowing P, T0, b,
R$_p$/\rstar, K and $\gamma$

To estimate the error on each fitted parameter a hybrid-MCMC approach
was used similar to Ford (2005).  The asteroseismic value of \rhostar\
and its statical error was adopted as a prior.  A
Gaussian Gibbs sampler was used to identify new jump values of test
parameter chains.  The width of the Gaussian sampler was initial
determined by the error estimates from the best-fit model.  After
500-chains where generated the chain success rate was examined and the
Gaussian width was rescaled using Equation 8 Gregory (2011).
This process was repeated after the
generation of each 500 chains until the success rate for each
parameter was between 22 and 28\%.  At which point the Gaussian width
was held fixed.

To handle the large correlation between the model parameters a hybrid
MCMC algorithm was adopted based on Gregory (2011).  The routine works
by randomly using a Gibbs sampler or a buffer of previously computed
chain points to generate proposals to jump to a new location in the
parameter space. The addition of the buffer allows for a calculation
of vectorized jumps that allow for efficient sampling of highly
correlated parameter space.  After the widths Gibbs sampler stabilized,
200,000 chains where generated.  The process was repeated 3 additional
times to test for convergence via a Gelman Rubin test (Gelman \& Rubin 1992).

The 4 chain sets where combined and used to calculate the median,
standard deviation and $1\sigma$ bounds of the parameter distribution
centered on the median value.  Adopting the asteroseismic errors on
the stellar mass and radius we computed the planetary radius ($R_p$),
inclination angle (i) and semi-major axis (a) and also the scaled
semi-major axis (a/\rstar) and transit-duration ($T_{dur}$) from the
model parameter distributions and report all values in Table 5.

\section{Addressing Blend Scenarios}

The lack of a clear Doppler detection needed for dynamical
confirmation of the nature of the transit signals in \hd\ requires us
to address the possibility that they are the result of contamination
of the light of the target by an eclipsing binary falling within the
photometric aperture (``blend''). The eclipsing binary may be either
in the background or foreground, or at the same distance as the target
in a physically associated configuration (hierarchical triple).
Furthermore, the object producing the eclipses may be either a star or
a planet.

We explore the wide variety of possible false positive scenarios using
the \blender\ technique (Torres et al. 2004, 2011; Freesin et al. 2011), 
which generates synthetic light curves for a large number of blend
configurations and compares them with the \kepler\ photometry in a
$\chi^2$ sense.  The parameters considered for these blends include
the masses (or spectral types) of the two eclipsing objects (or the
size of the one producing the eclipses, if a planet), the relative
distance between the binary and the target, the impact parameter, and
the eccentricity and orientation of the orbit of the binary, which can
affect the duration of the events.  Our simulations explore broad
ranges in each of these parameters, with the eccentricities for
planetary orbits limited to the maximum value recorded for known
transiting systems with periods as short as that of \hd\ (we adopted a
conservative limit of $e < 0.4$; see {\tt http://exoplanet.eu/}), and
eccentricities for eclipsing binaries limited to $e < 0.1$
(Raghavan et al. 2010).  Scenarios that give significantly worse fits
than a true transit model fit (at the 3-$\sigma$ level) are considered
to be rejected. While this rejection reduces the space of parameters
for viable blends considerably, it does not eliminate all possible
blends.  Constraints from follow-up observations described previously
(such as high-resolution imaging and spectroscopy) as well as
multi-band photometry available for the target allow us to rule out
additional areas of parameter space (see Figure~13). We then estimate
the {\it a priori} likelihood of the remaining blends in the manner
described in the next section.  To obtain a Bayesian estimate of the
probability that the transit events are due to a bona-fide planet, we
must compare the {\it a priori} likelihood of such a planet and of a
false positive (odds ratio).  We consider the candidate to be
statistically ``validated'' if the likelihood of a planet is several
orders of magnitude greater than that of a blend.\footnote{In the
context of this paper we reserve the term ``confirmation'' for the
unambiguous detection of the gravitational influence of the planet on
its host star (e.g., the Doppler signal) to establish the planetary
nature of the candidate; when this is not possible, as in the present
case, we speak of ``validation'', which involves an estimate of the
false alarm probability.} For full details on the \blender\ procedure,
we refer the reader to the references cited above.  Examples of other
\kepler\ candidates validated in this way include Kepler-9\,d
(Torres et al. 2011), Kepler-10\,c (Fressin et al.  2011), Kepler-11\,g
(Lissauer et al. 2011), Kepler-18\,b (Cochran et al 2011), and Kepler-19\,b
(Ballard et al. 2011).

\subsection{Background blends}
\label{sec:bp}

We examined first the case of background eclipsing binaries composed
of two stars.  Our detailed simulations with \blender\ indicate that
false positives of this kind are not able to match the observed shape
of the transit well enough (either in depth, duration of
ingress/egress, or total duration), or else they feature significant
ellipsoidal variations out of transit that are not seen in the
\kepler\ photometry. The best-fitting scenario of this kind gives a
match to the observations that is worse than that of a true transiting
planet model at the 6$\sigma$ level, which we consider unacceptable.
We also find that blends involving evolved stars (giants or subgiants)
orbited by a smaller star are easily ruled out, as well as those with
a main-sequence star eclipsed by a white dwarf. In both cases the
companion induces strong curvature out of eclipse due to the short
orbital period (2.78 days), and for giants the large stellar radius
additionally requires a grazing ``V''-shaped transit to match the
observed duration.

When the object producing the eclipses is a planet rather than a star,
ellipsoidal variations are negligible, and the shape of the eclipses
(further attenuated by the light of the target) can more easily match
the observed shape for a large range of properties of the stars and
planets involved. An illustration of the constraints provided by
\blender\ for false positives of this kind is shown in
Figure~\ref{fig:blender_bp}.  Following the \blender\ nomenclature we
refer to the target star as the ``primary'', and to the components of
the eclipsing pair as the ``secondary'' and ``tertiary'' (in this case
a planet).  The figure shows the $\chi^2$ landscape (goodness of fit
compared to a true transiting planet model) projected onto two of the
dimensions of parameter space, corresponding to the mass of the
secondary on the horizontal axis and the relative distance between the
primary and the binary on the vertical axis. The latter is cast for
convenience here in terms of the difference in distance modulus. The
colored regions represent contours of equal goodness of fit, with the
3-$\sigma$ contour indicated in white. Blends inside this contour give
acceptable fits to the \kepler\ photometry, and are considered viable.
They involve stars that can be up to 7 magnitudes fainter than the
target in the \kepler\ passband (as indicated by the dashed green line
in the figure, corresponding to background stars with $\Delta K\!p =
7$), and that are transited by a planet of the right size to produce
the measured signal. Also indicated are other constraints that rule
out portions of parameter space otherwise allowed by \blender.  The
blue hatched region represents blends that have overall colors for the
combined light as predicted by \blender\ that are either too red
(left) or too blue (right edge) compared to the measured color of the
target ($r\!-\!K_s = 1.314 \pm 0.035$, adopted from the
KIC; Brown et al. 2011), at the 3-$\sigma$ level. The green hatched area
represents blends that are bright enough (up to 4 mag fainter than the
target) to have been detected in our high-resolution spectroscopy as a
second set of lines (see Sect.~4). With these observational
constraints the pool of false positives of this kind is significantly
reduced, but many remain. We describe in Sect.~\ref{sec:validation}
how we assess their frequency.

\subsection{Blends involving physically associated stars}
\label{sec:htp}

Hierarchical triple configurations in which the eclipsing object
(tertiary) is a star are easily ruled out by \blender, as these
configurations invariably lead to the wrong shape for a
transit. However, stars physically associated with the target that are
orbited by a planet of the appropriate size can still mimic the light
curve well when accounting for dilution from the brighter star \hd.
The $\chi^2$ map for this type of blend is seen in
Figure~\ref{fig:blender_htp}. In this case the color of the blend is
not a strong discriminant, as all of these false positives are
predicted to have $r\!-\!K_s$ indices similar to that of the target
itself.  The expected brightness of the companion stars, though, is
such that most would have been detected spectroscopically ($\Delta
K\!p \leq 4$; green hatched exclusion region), unless their RV
compared to the target is small enough that their spectral lines are
blended with those of the main star. Based on our spectroscopic
observations we estimate conservatively that we would miss such
companions if they had radial velocities within $\sim$15\,\kms\ of the
RV of the target.  These blends are not eliminated by any other
observational constraint; we estimate their frequency below.

\subsection{Validation of Kepler-21\,b}
\label{sec:validation}

With the constraints on false positives afforded by the combination of
\blender\ and other follow-up observations, we may estimate the {\it a
priori} likelihood of a blend following a procedure analogous to that
explained by Fressin et al. (2001).

For blends involving background stars transited by a planet, this
frequency will depend on the density of background objects near the
target, the area around the target within which such stars would go
undetected, and the rate of occurrence of planets of the appropriate
size transiting those stars. We perform these calculations in
half-magnitude bins, with the following ingredients: \emph{a}) the
Galactic structure models of Robin et al. (2003)  to estimate the number
density of stars per square degree, subject to the mass limits allowed
by \blender; \emph{b}) results from our adaptive optics observations
to estimate the maximum angular separation ($\rho_{\rm max}$) at which
companions would be missed, as a function of magnitude difference
relative to the target ($K\!p = 8.224$) properly converted to the
$K\!p$ band, as described in the Appendix; and \emph{c}) the overall
frequency of suitable transiting planets that can mimic the signal.
The size range for these planets, as determined in our \blender\
simulations, is 0.38--2.0\,$R_{\rm Jup}$.  To estimate the frequency
of such planets we make use of the list of 1235 planet candidates
released by the \kepler\ Mission (Borucki et al. (2011), based on the
first four months of observation by the spacecraft. While these
objects have not all yet been confirmed because follow-up is still in
progress, the false positive rate is expected to be relatively low
\citep[typically less than 10\%; see][]{Morton:11} and will not affect
our results significantly. We therefore assume that all of them
represent true planets, and that the census of Borucki et al. (2011) is
complete for objects of this size (see below).  The estimated
frequency of these planets in the allowed radius range is $f_{\rm
planet} \approx 0.19$\%.

The results of our calculation for the frequency of blends involving
background stars is presented in Table~\ref{tab:blendfreq}.  Columns 1
and 2 give the magnitude range for background stars and the magnitude
difference compared to the target; column 3 lists the range of allowed
masses for the stars, based on our \blender\ simulations (see
Figure~\ref{fig:blender_bp}); columns 4 and 5 list the mean star
densities and $\rho_{\rm max}$, respectively, and column 6 gives the
number of background stars we cannot detect, and is the result of
multiplying column 4 by the area implied by $\rho_{\rm max}$. Finally,
the product of column 6 and the transiting planet frequency of 0.19\%
leads to the blend frequencies in column 7.  The sum of these
frequencies is given at the bottom under ``Totals'', and is $8.0
\times 10^{-7}$.

For blends involving physically associated stars with RVs within
15\,\kms\ of the target, which would go unnoticed in our spectroscopic
observations, we estimate the frequency through a Monte Carlo
experiment. We simulate companion stars in randomly chosen orbits
around the target, and randomly assign them transiting planets in the
appropriate radius range as a function of the secondary mass (see
Figure~\ref{fig:blender_htp}), according to their estimated
frequencies from Borucki et al. (2011).  We then determine what fraction
of these stars would be missed because of projected angular
separations below the $0\farcs05$ detection threshold from our speckle
observations, velocity differences relative to the target under
15\,\kms, or because they would induce a drift in the RV of the target
that is undetectable in our Keck observations (i.e., smaller than $\pm
10$\,\ms\ over a period of 82 days; see Sect.~4).  Binary orbital
periods, eccentricities, and mass ratios were drawn randomly from the
distributions presented by Raghavan et al. (2010), and the mass ratios
used in combination with our estimate of the mass of \hd\ to infer the
mass of the physical companions. We adopt an overall binary frequency
of 34\% from the same source.

Based on these simulations we obtain a frequency for this type of
false positive of $1.17 \times 10^{-6}$.  However, as seen in
Figure~\ref{fig:blender_htp}, planets involved in blends with
physically associated stars can be considerably smaller
($\sim$0.12--0.18\,$R_{\rm Jup}$) than those involved in background
blends (0.38--2.0\,$R_{\rm Jup}$), so we must consider the potential
incompleteness of the census of Borucki et al. (2011) at the smaller
planet sizes.  To estimate this we performed Monte Carlo simulations
in which we calculated the signal-to-noise ratio for each of the
\kepler\ targets that would be produced by a central transit of a
planet with the period of \hd\ and with a given radius in the allowed
range.  Adopting the \kepler\ detection threshold of 7 for the
signal-to-noise ratio Jenkins et al. (2010), we determined the fraction
of stars for which such a planet could have been detected during the
four months in which that sample was observed.  We have assumed that
the signal-to-noise ratio increases with the square root of the
transit duration and with the square root of the number of transits,
and that the data were taken in a continuous fashion (except for gaps
between quarters). In this way we obtained a completeness fraction of
about 65\%, although this may be slightly optimistic given that some
transits could have been missed due to additional interruptions in the
data flow for attitude corrections and safe mode events. This brings
the frequency of hierarchical triple blends to $1.8 \times 10^{-6}$.

The total blend frequency is then the sum of the two contributions
(background stars and physically associated stars with transiting
planets), which is $8 \times 10^{-7} + 1.8 \times 10^{-6} = 2.6 \times
10^{-6}$.

Finally, following the Bayesian approach outlined earlier, we require
also an estimate of the likelihood of a true planet around \hd\
(``planet prior'') to assess whether it is sufficiently larger than
the likelihood of a blend, in order to validate the candidate.  To
estimate the planet prior we may appeal once again to the catalog of
1235 candidates from Borucki et al. (2011), which contains 99 systems with
planetary radii within 3$\sigma$ of the measured value for \hd\ ($R_p
= 1.64 \pm 0.04\,R_{\earth}$). The 3-$\sigma$ limit used here is for
consistency with a similar criterion adopted above in \blender. Given
the total number of 156,453 \kepler\ targets from which the 1235
candidates were drawn, we obtain a planet frequency of $99/156,\!453 =
6.3 \times 10^{-4}$. Applying the same incompleteness factor described
above, which holds also for the radius of this candidate, we arrive at
a corrected planet prior of $9.7 \times 10^{-4}$.  This conservative figure is
$\sim$370 times larger than the blend frequency, which we consider
sufficient to validate the planet around \hd\ to a high degree of
confidence. We note that this odds ratio is a lower limit, as we have
been conservative in several of our assumptions. In particular, for
computing the frequency of planets transiting background stars
(Table~\ref{tab:blendfreq}) we have included objects with sizes
anywhere between the minimum and maximum planet radius allowed by
\blender\ for stars of all spectral types (0.38--2.0\,$R_{\rm Jup}$),
whereas the planet size range for secondaries of a given mass is
considerably smaller.  This would reduce the frequency of this type of
false positive, strengthening our conclusion.

\section{Limits to the Density and Mass for Kepler-21b}

To determine a statistically firm upper limit to the planet mass, we carried out an MCMC analysis of the 
Keck radial velocities with a Keplerian
model for the planet's orbit.
The resulting 2-sigma upper limit to the mass yielded the Keplerian model shown in Figures 9 \& 10 and gave the
following upper limits:  RV amplitude of $K<$3.9 m s$^{-1}$; a planet mass of $M=10.4$ $M_{\rm Earth}$ (2$\sigma$),
and a corresponding density of $\rho<12.9$ g cm$^{-3}$. 
This upper limit to the density of 12.9$g cm^{-3}$ is so high that the planet could be (compressed) solid or composed of
admixtures of rocky, water, and gas in various amounts, unconstrained by this large upper limit to density.  The 1-sigma upper
limit to density is 7.4$g cm^{-3}$, still consistent with all types of interior compositions and would 
yield a planet mass of $\sim$5.9 M$_{\rm Earth}$. 

If Kepler-21b contains a large rocky core, the
high pressure inside such a massive planet would cause the silicate mantle 
minerals to compress to dense phases of post-perovskite; the iron core is 
also at higher density than inside Earth (Valencia et al. 2007).
However, Kepler-21b could also have a small rocky core, be mostly gas and not be nearly as massive.
The maximum core fraction expected for rocky planets of this radius 
corresponds to a planet with mass of 10.0 M$_{\rm Earth}$ and a mean density of 12.5 
g/cc (see mantle stripping simulations by Marcus et al. 2010) with a corresponding RV semi-amplitude of 
2.3 m/sec, still below our detection limit. If Kepler-21b
is a water planet with low silicate-to-iron ratio and 50\% water 
by mass, its mass would be merely 2.2 M$_{\rm Earth}$, similar to that of 
Kepler-11f, but at mean density of 2.7 g/cc.
The measured radial velocities provide neither a
confirmation nor a robust limit ($\sim$10 Earth-masses) on the mass of Kepler-21b but
suggests an upper limit near that of the maximum rocky core fraction theoretically allowed.
The radial velocities certainly rule out higher mass
companions (additional planets or stellar companions) with orbital periods in the period range 
of up to
approximately 200 days as any RV trend caused by such a companion would be apparent 
in Figs. 9 \& 10.

\section{Conclusion}

{\it Kepler} photometry of the bright star HD 179070 reveals a small periodic transit-like signal
consistent with a 1.6 R$_{\rm Earth}$ exoplanet. The transit signal repeats every 2.8 days and the complete
phased light curve shows all of the events to be consistent in phase, amplitude, and duration.
Analysis of the {\it Kepler} image data and difference images are well matched by model fits. 
Detailed point response function (PRF) models conclude that the source of the transit
event is centered on or near to the center of HD 179070 itself. Furthermore, these models
show that many faint background eclipsing binary scenarios, capable of blending light with that from HD 179070 to
produce the transit-like event, can be eliminated. 

High resolution ground-based optical speckle imaging reveals no nearby companion star to within 5 magnitudes of HD179070
itself. Near-IR AO observation, however, reveal a faint companion star 0.75" away and $\sim$4 magnitudes fainter in $K$.
Using a color transformation, this star  is expected to be R$\sim$14.2, just below the detection limit of the speckle
results. Spectroscopic observations also confirm that no bright star is present near 
HD 179070 (within 0.5"). 
Asteroseismology was performed for HD 179070 using the {\it Kepler} light curves. Adopting the
spectroscopically determined values for T$_{\rm eff}$, log g, and [Fe/H], the mass,  radius, and age of HD 179070
were well determined. 

Putting all of the above observations and models together, we conclude that the cause of the 
periodic transit event is indeed a small 1.6 R$_{\rm Earth}$ exoplanet orbiting the subgiant star HD 179070.
Transit models were fit to the highly precise {\it Kepler} light curve data
revealing that the exoplanet orbits every 2.78 days at an
inclination of 82.5 degrees. The exoplanet has an equilibrium temperature near 1900K and is located 0.04 AU
from its host star. 
Kepler-21b has been validated by detailed modeling of blend scenarios as a true exoplanet at greater than 99.7\% confidence. 
We can only determine an upper mass limit for the exoplanet, $\sim$10 M$_{\rm Earth}$, resulting in a upper limit to
the mean density of $\sim$13 g cm$^{-3}$. 

{\it Kepler} continues to monitor HD 179070 and will eventually build up higher S/N phased transit light
curves. These long term observations may allow other planets within this same system may be directly 
detected or detected via transit timing variations.
Given the brightness of HD 179070, it is likely that continued radial velocity monitoring will take place
with Keck or other current or planned radial velocity instruments. Given a consistent level of instrumental
precision, the observed stellar jitter will slowly be averaged out and velocity signals from this or 
other planets orbiting the host star may be detected. 
Finally, using a technique such as described in Schuler et al. (2011) high-resolution, high signal-to-noise
echelle spectroscopy will provide detailed metal abundance values of 
the host star's atmosphere which may hold clues as to the formation, or not, of planetary bodies.

\acknowledgments{
We thank John Johnson for use of some of his Keck Time.
This research has made use of the NASA/IPAC/NExScI Star
and Exoplanet Database, which is operated by the Jet Propulsion Laboratory,
California Institute of Technology, under contract with the National
Aeronautics and Space Administration.
The authors would like to thank the
{\it Kepler} Science Office and the Science Operations Center personal
for their
dedicated effort to the mission and  for providing us access to the science
office data products. The ground-based observations reported on herein were
obtained at  Kitt Peak National Observatory, National Optical Astronomy
Observatory, which is operated by the Association of Universities for
Research in Astronomy (AURA) under cooperative agreement with the National
Science Foundation. 
{\it Kepler} was
selected as the 10th mission of the Discovery Program. Funding for this
mission is provided by NASA.}

\section{Appexdix A: Transformation of Infrared Colors}

To understand the contribution of the faint infrared companion to the light
curve in the Kepler bandpass, we need to convert the measured infrared
color ($J-K^\prime$) to a Kepler magnitude ($Kp$). To do this, we have 
derived a color-color relationship ($Kp - Ks$ vs. $J-Ks$) utilizing the
Kepler targets from Q1 public release and the  photometry from the Kepler
Input Catalog (KIC; Brown et al. 2011).  Separating the KIC into dwarfs and
giants as described by Ciardi et al. (2011), we have fitted the color-color
relationship with a 5th-order polynomial for the dwarfs and a 3rd-order
polynomial for the giants (see Figures~\ref{fig:ccplot-dwarf} and
\ref{fig:ccplot-giant}).

The dwarf and giant color-color relationships were determined separately
from the Kepler magnitude and 2MASS  magnitudes of 126092 dwarfs within the
color range of $-0.2 \leq J-Ks \leq 1.0$ mag and 17129 giants within  the
color range of the $-0.2 \leq J-Ks \leq 1.2$ mag.  The resulting polynomial
coefficients from the least squares fits for the dwarfs and giants, 
respectively, are
\[
 {\rm Dwarfs:\ } Kp - Ks = 0.314377 + 3.85667x + 3.176111x^2 - 25.3126x^3 +
 40.7221x^4  - 19.2112x^5
\]
\[
 {\rm Giants:\ } Kp - Ks = 0.42443603 + 3.7937617x - 2.3267277x^2 + 1.4602553x^3
\]
\[
{\rm where\ } x = J - Ks.
\]

The fits and the residuals are shown in  Figures~\ref{fig:ccplot-dwarf} and
\ref{fig:ccplot-giant}; the residuals for both the dwarfs and giants are 
well-characterized by gaussian distributions with means and  widths of
$\langle Kp_{J-Ks} 
- Kp_{true}\rangle = -0.005 \pm 0.083$ mag and
$\langle Kp_{J-Ks} 
- Kp_{true}\rangle = -0.002 \pm 0.065$ mag
for the dwarfs and giants, respectively.  The
uncertainties in the derived Kepler magnitudes ($Kp$) are dominated by the
physical widths of the color-color relationships.

The real apparent photometry of the infrared companion was determined from 
the 2MASS photometry of the primary target ($J=7.229\pm0.032$ mag,
$Ks=6.945 \pm 0.018$ mag) which is a blend of the two sources.  The above
color-color relationships were determined using the $Ks$ filter, but the
observations were taken in the $K^\prime$ filter which has a slightly
shorter central wavelength (2.148 $\mu$m vs. 2.124 $\mu$m).  Typically, the
$Ks$ and $K^\prime$ filters yield magnitudes which are within $0.02-0.03$
magnitudes of each other and have zero-point flux densities within 2\%
(AB magnitudes = 1.86 and 1.84 for $Ks$ and $K^\prime$, respectively;
Tokunaga \& Vacca 2005).  Given the quality of the weather and resulting photometry, the
width of the color-color relationships, and lack of an $H$-band observation
to aid in the transformation of the $K^\prime$ observations, we have
equated $K^\prime$ to $Ks$ in these calculations and propagated an
additional uncertainty of 0.03 mag in the derivation of the $J-Ks$ color
and the Kepler magnitude ($Kp$).

Deblending the infrared photometry results in the following infrared
magnitudes for the primary target and faint companion of $J=7.24 \pm 0.07$ 
and $11.94 \pm 0.07$ mag,respectively, and $Ks = 6.97 \pm 0.07$ and $Ks =
10.92  \pm 0.07$ mag, where we have propagated the uncertainties of the
faint companion on to the photometry of both stars.  The infrared colors of
the faint companion is $J-Ks = 1.0\pm0.1$ mag, which corresponds to a
$Kp-Ks$ color of $Kp-Ks = 3.54 \pm 0.14$ mag if the faint star is a dwarf
and a color of $Kp-Ks = 3.35 \pm 0.14$ mag if the faint star is a giant. 

Applying the deblended $Ks$ magnitude of the companion ($Ks = 10.92 \pm
0.07$ mag), we derive a magnitude for the faint star in the  Kepler
bandpass for the dwarf- and giant-star relationships of $Kp = 14.5 \pm 0.2$
mag and $Kp = 14.3 \pm 0.2$ mag, respectively.  The  companion is fainter
than the primary target, in the Kepler bandpass, by $\Delta Kp = 6.3$ mag
if the star is dwarf and $\Delta Kp = 6.1$ mag if the star is a giant. 
Note the primary star dominates the photometry in the Kepler aperture;
after deblending the Kepler magnitude of the star changes from $Kp = 8.224$
mag to $Kp = 8.227$ mag if the companion is a dwarf or to $Kp = 8.228$ mag
if the companion is a giant.

The above relationships only work if both $J$ and $Ks$ magnitudes are known,
but often only one of the filters is available.  Being able to convert a
single $J$ and $Ks$ magnitude into an expected $Kp$ is extremely useful
-- particularly,  for determining sensitivity limits for the AO imaging.
Towards this end we have utilized the Kepler Input Catalog to determine the
expected $Kp - J$ and $Kp - Ks$ colors for a given $J$ or $Ks$ 
magnitude.  

Histograms of the $Kp - J$ and $Kp - Ks$ colors are shown in Figure
\ref{fig:kepmag_color} where the median and mode of the color are marked;
the spread in color, as described by the dispersion of the colors, is
fairly large.  The medians, modes (both of which are delineated in the
histograms), and dispersions of the colors are  $Kp - J = 1.477$,
$1.275$, and $0.626$ mag and $Kp - Ks = 2.139$, $1.775$, and $0.803$
mag. 

It is not unexpected that the measured median colors would be dependent
upon the real apparent infrared magnitude; as the photometry becomes more
sensitive to fainter and fainter sources, more intrinsically fainter (and
redder) sources should contribute more significantly to the color
distribution.  To explore this effect, we have computed the median color
($Kp - J$ and $Kp - Ks$) as a function of the real apparent 
infrared magnitude (see Fig~\ref{fig:kepmag_color}).  The dispersion per
bin is fairly large ($0.4 - 0.9$ mag), but the colors show smooth systematic
trends as a function of magnitude with a range of $0.5 - 0.6$ mag.  

We have characterized these curves with 5th-order polynomials and have done
a linear extrapolation for magnitudes fainter than the data range:
\[
 Kp - J = -398.04666 + 149.08127J - 21.952130J^2 + 1.5968619J^3 -
 0.057478947J^4  + 0.00082033223J^5 
\]
for $(10 < J < 16.7\ {\rm mag})$ 
\[
 Kp - J = 0.1918 + 0.08156J 
 \]
for $(J > 16.7\ {\rm mag})$ and 
\[
 Kp - Ks = -643.05169 + 246.00603Ks - 37.136501Ks^2 + 2.7802622Ks^3 -
 0.10349091Ks^4  + 0.0015364343Ks^5 \] 
for $(10 < Ks < 15.4\ {\rm mag})$ 
\[
 Kp - Ks = -2.7284 + 0.3311Ks 
\]
 for $(Ks > 15.4\ {\rm mag})$. 

The trends seen in the color vs. magnitude relationships are not
unexpected. At the brighter magnitudes, the distribution of stars is
dominated by  infrared bright stars (i.e. giants) and thus, are dominated
by relatively red stars.  As the magnitude limit is increased, the dwarf
stars begin to  contribute to the color distribute starting with the bluer
(more luminous) stars and the median colors become bluer.  As the magnitude
limits are pushed even further, the intrinsically fainter (i.e., red) dwarf
stars begin to dominate the sample, and the median colors become
increasingly red as the magnitude limit is increased.  

\newpage

\begin{deluxetable}{ccccc}
\tablecolumns{5}
\tablewidth{4in}
\tablecaption{Approximate Radial Source Sensitivity \label{tab:keckAO}} 
\tablehead{
\colhead{Distance} & \colhead{Distance} & \colhead{$\Delta J$} & 
\colhead{Distance} & \colhead{$\Delta Ks$} \\
\colhead{(FWHM)} & \colhead{$(\prime\prime)$} & \colhead{(mag)} & 
\colhead{$(\prime\prime)$} & \colhead{(mag)}
}
\startdata
1 & 0.07  &  1.0 & 0.09  &  1.5 \\
2 & 0.14  &  1.5 & 0.18  &  2.0 \\
3 & 0.21  &  2.0 & 0.27  &  2.5 \\
4 & 0.28  &  2.5 & 0.36  &  3.0 \\
5 & 0.35  &  3.0 & 0.45  &  3.5 \\
6 & 0.42  &  3.5 & 0.54  &  4.0 \\
7 & 0.49  &  4.0 & 0.63  &  4.5 \\
8 & 0.56  &  4.5 & 0.72  &  5.0 \\
9 & 0.63  &  5.0 & 0.81  &  5.5 \\ 
10& 0.70  &  5.5 & 0.90  &  6.0 \\
11& 0.77  &  6.0 & 0.99  &  6.5 
\enddata
\end{deluxetable}

\begin{deluxetable}{cccc}
\tablecolumns{4} \tablewidth{0pc}
\tablecaption{{\it Kepler} image centroid offsets in pixels for HD 17907} 
\tablehead{ \colhead{Quarter\tablenotemark{a}}& \colhead{Row offset}& \colhead{Column offset}& \colhead{Offset distance}}
\startdata
1&  $2.09 \pm 0.02$&  $1.23 \pm 0.03$&  $2.43 \pm 0.02$\\
3&  $-0.44 \pm 0.03$&  $1.18 \pm 0.07$&  $1.26 \pm 0.07$\\
4&  $0.18 \pm 0.07$& $-0.35 \pm 0.07$& $0.39 \pm 0.07$\\
5& $-0.30 \pm 0.02$& $0.35 \pm 0.03$& $0.46 \pm 0.03$\\
\enddata
\tablenotetext{a}{Quarter 1 was short (1 month) and provides less reliable centriod measurements. Quarter 2 data was not 
included due to its excess noise.}
\label{tab:centroids}
\end{deluxetable}

\begin{deluxetable}{lcccc}
\tablenum{2}
\tablecolumns{5} 
\tablewidth{0pc} 
\tablecaption{Spectral Analysis of HD 179070}
\tablehead{ 
\colhead{Source} & 
\colhead{T$_{eff}$ (K)} & 
\colhead{log g (cgs)} & 
\colhead{[Fe/H]} & 
\colhead{Vsini (km/sec)} 
}
\startdata
N(2004)\tablenotemark{a} & 6137 & --- & -0.015 & --- \\
KPNO 4-m & 6250$\pm$250 &  4.0$\pm$0.25 &  -0.15$\pm$0.15 & --- \\
TRES & 6250$\pm$125 &  4.0$\pm$0.25 &  0.0$\pm$0.25 &  8.0$\pm$1.0 \\
Keck HIRES &  6131$\pm$44 & 3.9$\pm$0.1 &  -0.05$\pm$0.1  & 7.5$\pm$1.0 \\
MZ(2010)\tablenotemark{b} & 6063$\pm$126 & 4.04$\pm$0.07 & -0.23$\pm$0.09 & $<$5 \\
MZ(2010)\tablenotemark{b} & 6145$\pm$65 & 4.15$\pm$0.10 & -0.15$\pm$0.06 & $<$5 \\
HB\tablenotemark{c} & --- & --- & -0.15 & --- \\
Adopted & 6131$\pm$44 & 4.0$\pm$0.1 & -0.15$\pm$0.06 & 7.75$\pm$1.0 \\
\enddata
\tablenotetext{a} {Nordstrom et al., 2004}
\tablenotetext{b} {Molenda-Zakowicz et al., 2010, two solutions listed}
\tablenotetext{c} {H. Bruntt, private communication}
\end{deluxetable}

\begin{deluxetable}{rrr}
\tablenum{3}
\tablecaption{Radial Velocities for HD 179070}
\label{}
\tablewidth{0pt}
\tablehead{
\colhead{BJD}         & \colhead{RV}     & \colhead{Unc.}  \\
\colhead{-2440000}   & \colhead{(m/sec)}  & \colhead{(m/sec)}  }
\startdata
      15439.938 &  2.61  & 2.1 \\
      15439.941 &  8.44  & 2.2 \\
      15439.943 &  3.99  & 2.2 \\
      15440.771 &  -0.53 &  2.2 \\
      15440.773 &  -0.73 &  2.0 \\
      15440.775 &  -6.18 &  2.2 \\
      15440.987 &  -3.99 &  2.3 \\
      15440.989 &  -5.95 &  2.3 \\
      15440.991 &  0.84  & 2.2 \\
      15455.826 &  -3.01 &  2.1 \\
      15455.828 &  -7.30 &  2.1 \\
      15455.830 &  2.74  & 2.0 \\
      15464.788 &  14.96 &  1.8 \\
      15464.790 &  10.45 &  1.8 \\
      15464.793 &  10.72 &  1.8 \\
      15465.866 &  14.74 &  1.8 \\
      15465.869 &  4.12  & 1.9 \\
      15465.872 &  7.47  & 1.9 \\
      15466.725 &  -0.56 &  1.9 \\
      15466.727 &  -3.96 &  2.0 \\
      15466.729 &  1.87  & 2.0 \\
      15467.848 &  -9.35 &  1.9 \\
      15467.849 &  -0.91 &  2.0 \\
      15467.851 &  3.14  & 2.1     \\                                        
      15468.714 &  -1.64 &  2.0 \\
      15468.716 &  2.15  & 1.9 \\
      15468.718 &  1.55  & 2.0 \\
      15469.753 &  -4.89 &  1.7 \\
      15469.755 &  -8.50 &  1.8 \\
      15469.758 &  1.13  & 1.8 \\
      15471.847 &  1.18  & 2.0 \\
      15471.849 &  -3.06 &  1.8 \\
      15471.852 &  -7.47 &  1.8 \\
      15486.819 &  -2.46 &  2.0 \\
      15486.822 &  0.56  & 2.2 \\
      15486.824 &  -2.85 &  2.0 \\
      15490.819 &  -3.67 &  2.0 \\
      15490.821 &  -5.84 &  2.1 \\
      15490.823 &  -7.85 &  1.9 \\
      15521.759 &  -8.33 &  2.2 \\
\enddata
\end{deluxetable}
\clearpage

\clearpage

\begin{deluxetable}{cccc}
\tablenum{4}
\tablecolumns{4} \tablewidth{0pc} 
\tablecaption{Estimated frequencies $\nu_{nl}$ of HD 179070 (in $\rm \mu
 Hz$).}
\tablehead{ \colhead{$n$}& \colhead{$l=0$}& \colhead{$l=1$}& \colhead{$l=2$}} 
\startdata
12&          ...      &          ...      &  $850.13 \pm 3.57$\\
13&  $855.05 \pm 2.09$&  $885.51 \pm 1.23$&  $907.92 \pm 4.93$\\
14&  $918.26 \pm 1.16$&  $946.67 \pm 1.28$&  $975.03 \pm 1.31$\\
15&  $979.45 \pm 0.25$& $1005.25 \pm 0.75$& $1034.88 \pm 1.70$\\
16& $1039.33 \pm 0.42$& $1064.76 \pm 0.72$& $1095.85 \pm 1.23$\\
17& $1098.37 \pm 0.86$& $1125.63 \pm 0.44$& $1155.74 \pm 1.46$\\
18& $1159.28 \pm 0.88$& $1187.41 \pm 0.58$& $1215.74 \pm 1.95$\\
19& $1221.45 \pm 1.03$& $1248.53 \pm 0.61$& $1279.21 \pm 1.69$\\
20& $1282.74 \pm 0.85$& $1308.73 \pm 0.71$& $1339.38 \pm 1.45$\\
21& $1341.48 \pm 0.56$& $1370.70 \pm 1.02$& $1399.88 \pm 2.19$\\
22& $1404.24 \pm 1.85$& $1432.05 \pm 1.72$&          ...      \\
\enddata
\label{tab:freqs}
\end{deluxetable}

\begin{deluxetable}{lccccc}
\tablenum{5}
\tablewidth{0in}
\tablecaption{Transit Model Parameters (e=0)\label{ta:model1}}
\tablehead{
\colhead{ } & \colhead{Bestfit\tablenotemark{\dag}} & \colhead{Median} & \colhead{Stdev} & 
\colhead{+1$\sigma$}  & \colhead{-1$\sigma$}
}
\startdata
Adopted Values & & & & & \\
\mstar\ (\msun)              &   1.340 &   -- &   0.010 &   0.010 &  -0.010\\
\rstar\ (\rsun)              &   1.860 &   -- &   0.020 &   0.020 &  -0.020\\
\logg$_\star$                &   4.0190 &   4.0196 &   0.0090 &   0.0087 &  -1.0694\\
\rhostar\ (g/cm$^3$)         &   0.2886 &   0.2891 &   0.0087 &   0.0077 &  -0.0102\\
\hline
Derived Values & & & & & \\
R$_p$ (\rjup)                &   0.1459 &   0.1456 &   0.0035 &   0.0034 &  -0.0038\\
P (days)                     &   2.785755 &   2.785755 &   0.000032 &   0.000031 &  -0.000034\\
i (deg)                      &  82.58 &  82.59 &   0.29 &   0.28 &  -0.31\\
T0\tablenotemark{\ddag}      &  193.8369 &  193.8368 &   0.0016 &   0.0016 &  -0.0016\\
a/\rstar\                    &   4.910 &   4.913 &   0.050 &   0.043 &  -0.058\\
R$_p$/\rstar\                &   0.00806 &   0.00804 &   0.00018 &   0.00018 &  -0.00019\\
b                            &   0.640 &   0.639 &   0.023 &   0.020 &  -0.028 \\
a (AU)                       &   0.042507 &   0.042509 &   0.000106 &   0.000098 &  -0.000119\\
T$_{dur}$ (h)                &   3.438666 &   3.438982 &   0.078588 &   0.070336 &  -0.091437\\
T$_{eq}$                     &  1956$\pm$297 & & & &  \\ 
\enddata
\tablenotetext{\dag}{\mstar\ and \rstar\ are fixed to asteroseismic values}
\tablenotetext{\ddag}{T0=BJD-2454900}
\end{deluxetable}

%
%

\begin{deluxetable}{ccccccc}
\tablenum{6}
\tabletypesize{\scriptsize}
\tablewidth{0pc}

\tablecaption{Blend frequency estimate for \hd\ for scenarios
involving background stars transited by a
planet.\label{tab:blendfreq}}

\tablehead{
\colhead{$K\!p$ range} &
\colhead{$\Delta K\!p$} &
\colhead{Stellar mass} &
\colhead{Stellar density} &
\colhead{$\rho_{\rm max}$} &
\colhead{Stars} &
\colhead{Blends\tablenotemark{a}} \\
\colhead{(mag)} &
\colhead{(mag)} &
\colhead{range ($M_{\odot}$)} &
\colhead{(per sq.\ deg)} &
\colhead{(\arcsec)} &
\colhead{($\times 10^{-6}$)} &
\colhead{($\times 10^{-6}$)} \\
\colhead{(1)} &
\colhead{(2)} &
\colhead{(3)} &
\colhead{(4)} &
\colhead{(5)} &
\colhead{(6)} &
\colhead{(7)}
}
\startdata
8.2--8.7    &  0.5 & \nodata   & \nodata&\nodata & \nodata& \nodata \\
8.7--9.2    &  1.0 & \nodata   & \nodata&\nodata & \nodata& \nodata \\
9.2--9.7    &  1.5 & \nodata   & \nodata&\nodata & \nodata& \nodata \\
9.7--10.2   &  2.0 & \nodata   & \nodata&\nodata & \nodata& \nodata \\
10.2--10.7  &  2.5 & \nodata   & \nodata&\nodata & \nodata& \nodata \\
10.7--11.2  &  3.0 & \nodata   & \nodata&\nodata & \nodata& \nodata \\
11.2--11.7  &  3.5 & \nodata   & \nodata&\nodata & \nodata& \nodata \\
11.7--12.2  &  4.0 & \nodata   & \nodata&\nodata & \nodata& \nodata\\
12.2--12.7  &  4.5 & 0.80--1.40  & 115   &  0.50  &  6.97  & 0.0132 \\
12.7--13.2  &  5.0 & 0.80--1.40  & 204   &  0.60  &  17.8  & 0.0338 \\
13.2--13.7  &  5.5 & 0.80--1.40  & 267   &  0.75  &  36.4  & 0.0692 \\
13.7--14.2  &  6.0 & 0.80--1.40  & 391   &  0.85  &  68.5  & 0.130 \\
14.2--14.7  &  6.5 & 0.80--1.28  & 438   &  0.95  &  95.8  & 0.182 \\
14.7--15.2  &  7.0 & 0.85--1.15  & 733   &  1.05  &  195.9  & 0.372 \\
15.2--15.7  &  7.5 & \nodata     & \nodata&\nodata & \nodata& \nodata \\
15.7--16.2  &  8.0 & \nodata     & \nodata&\nodata & \nodata& \nodata \\
\noalign{\vskip 6pt}
\multicolumn{2}{c}{Totals} & & 2148 &  & 421.4 & {\bf 0.800} \\
\noalign{\vskip 4pt}
\hline
\noalign{\vskip 4pt}
\multicolumn{7}{c}{Total blend frequency = $8.0 \times 10^{-7}$}
\enddata

\tablenotetext{a}{The range of radii allowed by \blender\ for the
planets involved in these blends is 0.38--2.0\,$R_{\rm Jup}$, and the
planet frequency used for the calculation is $f_{\rm planet}=0.19$\%
(see text).}

\tablecomments{Magnitude bins with no entries correspond to brightness
ranges in which \blender\ excludes all blends, or that are ruled out
by spectroscopic constraints.}

\end{deluxetable}

\clearpage

\begin{figure}
\includegraphics[angle=0,scale=0.6,keepaspectratio=true]{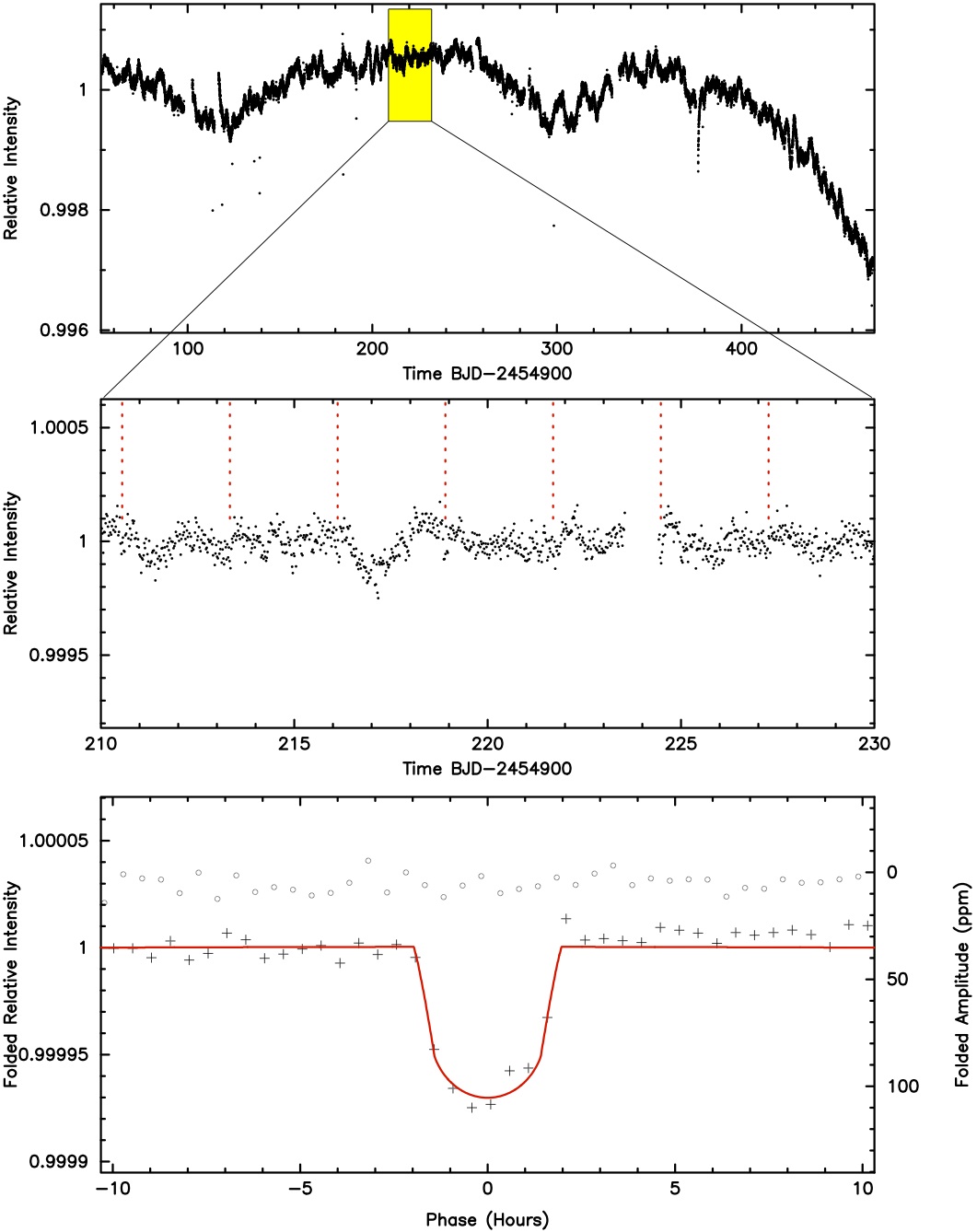}
\caption{{\it Kepler} light curve of HD 179070 covering Quarters 0 to 5. 
The top raw light curve covers 164 separate transit events for
the small exoplanet orbiting the star. 
The middle panel shows a typical normalized section from the full light curve in which transits 
are visible (positions marked with dotted lines).
The bottom panel shows the detrended, binned, and phase folded-data (see \S6) overplotted by our model fit (red line).
}
\end{figure}


\begin{figure}
\includegraphics[angle=0,scale=0.6,keepaspectratio=true]{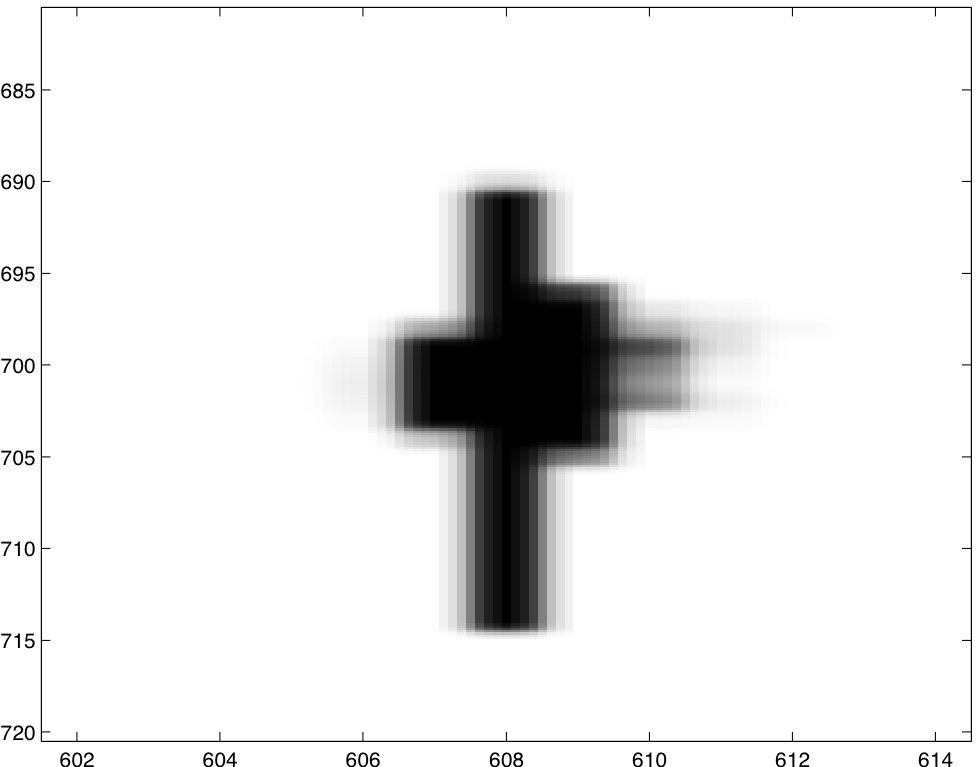}
\caption{A typical {\it Kepler} Quarter 5 pixel image of HD 179070 showing the saturation
spilled along columns and the non-saturated wings around the core.}
\label{fig:directImage}
\end{figure}

\begin{figure}
\includegraphics[angle=0,scale=0.6,keepaspectratio=true]{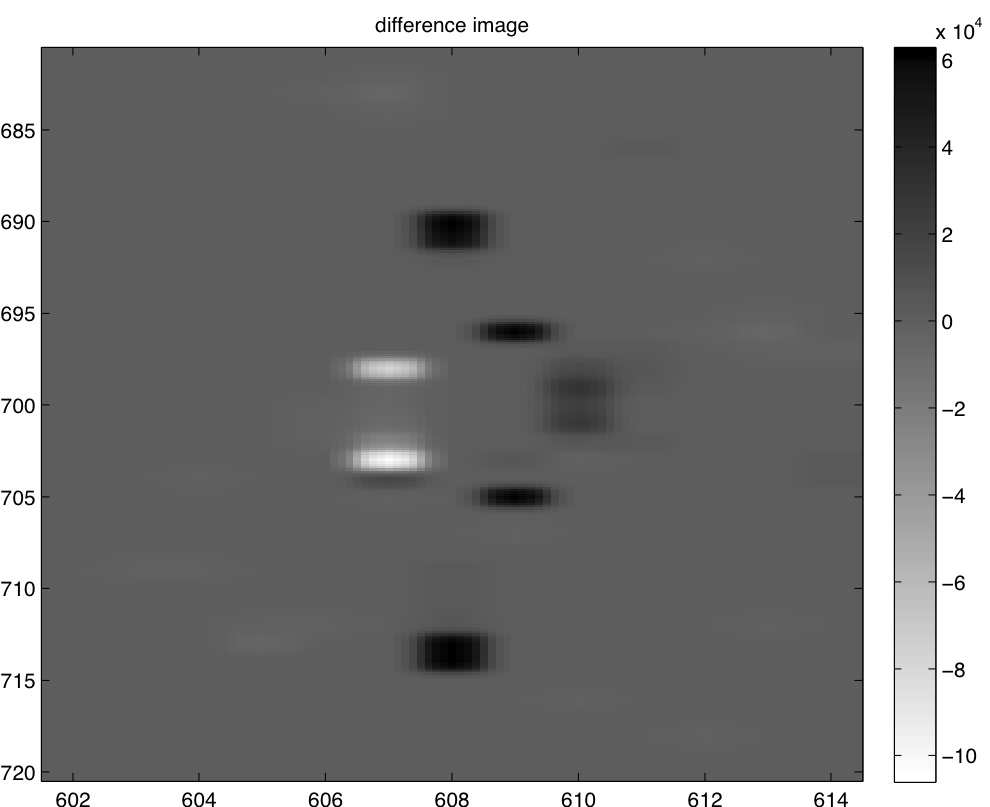}
\caption{{\it Kepler} Quarter 5 average difference image of HD 179070 showing the pixels that 
change during the transit event.}
\label{fig:diffImage}
\end{figure}

\begin{figure}
\includegraphics[angle=0,scale=0.6,keepaspectratio=true]{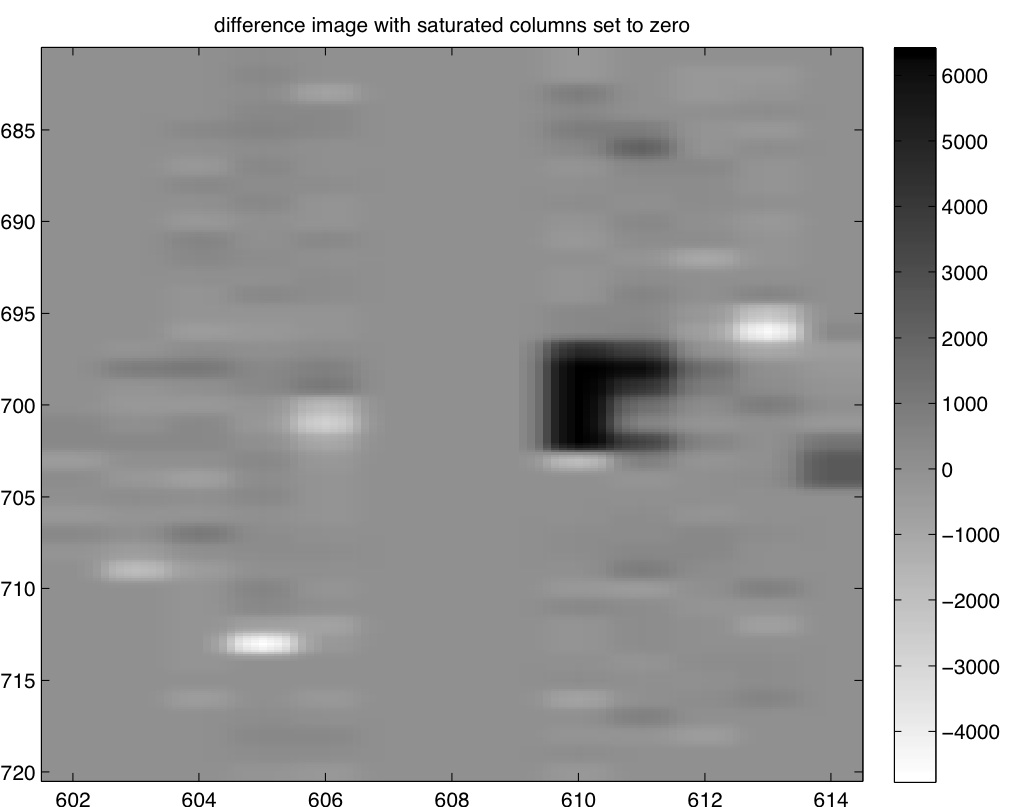}
\includegraphics[angle=0,scale=0.6,keepaspectratio=true]{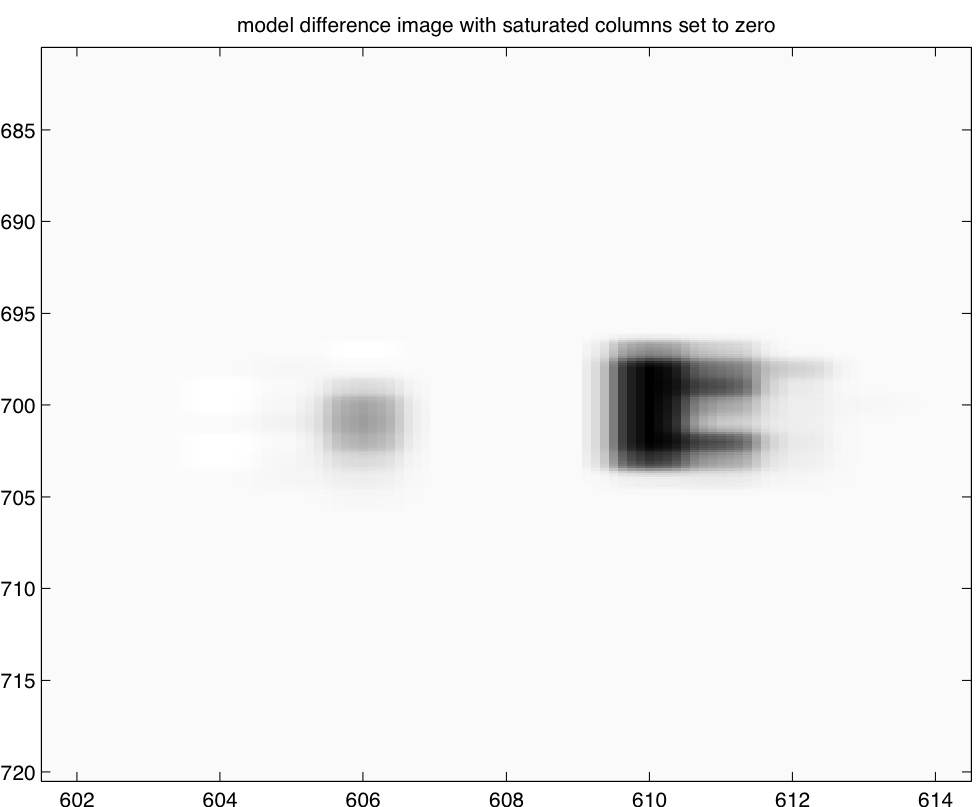}
\caption{Left: {\it Kepler} Quarter 5 average difference image with the saturated pixels set to zero.
Right: The corresponding PRF model difference image.}
\label{fig:noSatDiffImage}
\end{figure}

\begin{figure}
\includegraphics[angle=0,scale=0.6,keepaspectratio=true]{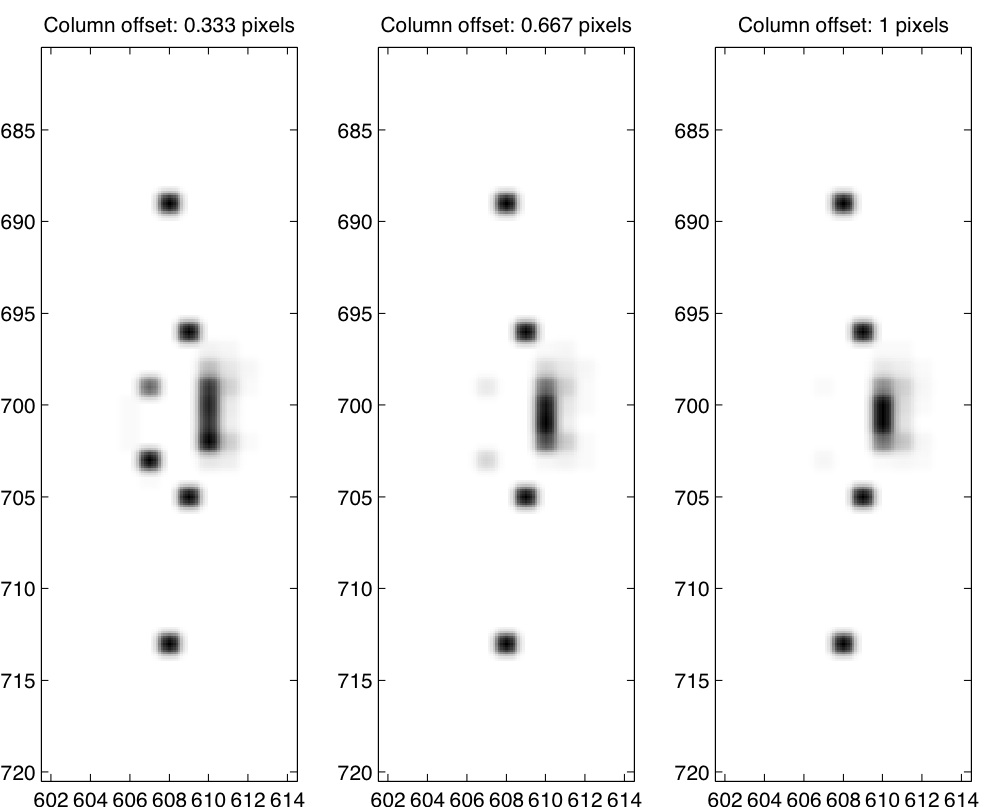}
\caption{Quarter 5 model difference images in which the modeled faint background eclipsing binary
star was offset in the column direction from the center location
of HD 179070. As the column offset increases (left to right) the difference signal in the left most saturated
pixels goes away. A similar result is seen for a column shift to the left.}
\label{fig:colOffsetDiffImage}
\end{figure}


\begin{figure}
\epsscale{0.8}
\includegraphics[angle=0,scale=0.6,keepaspectratio=true]{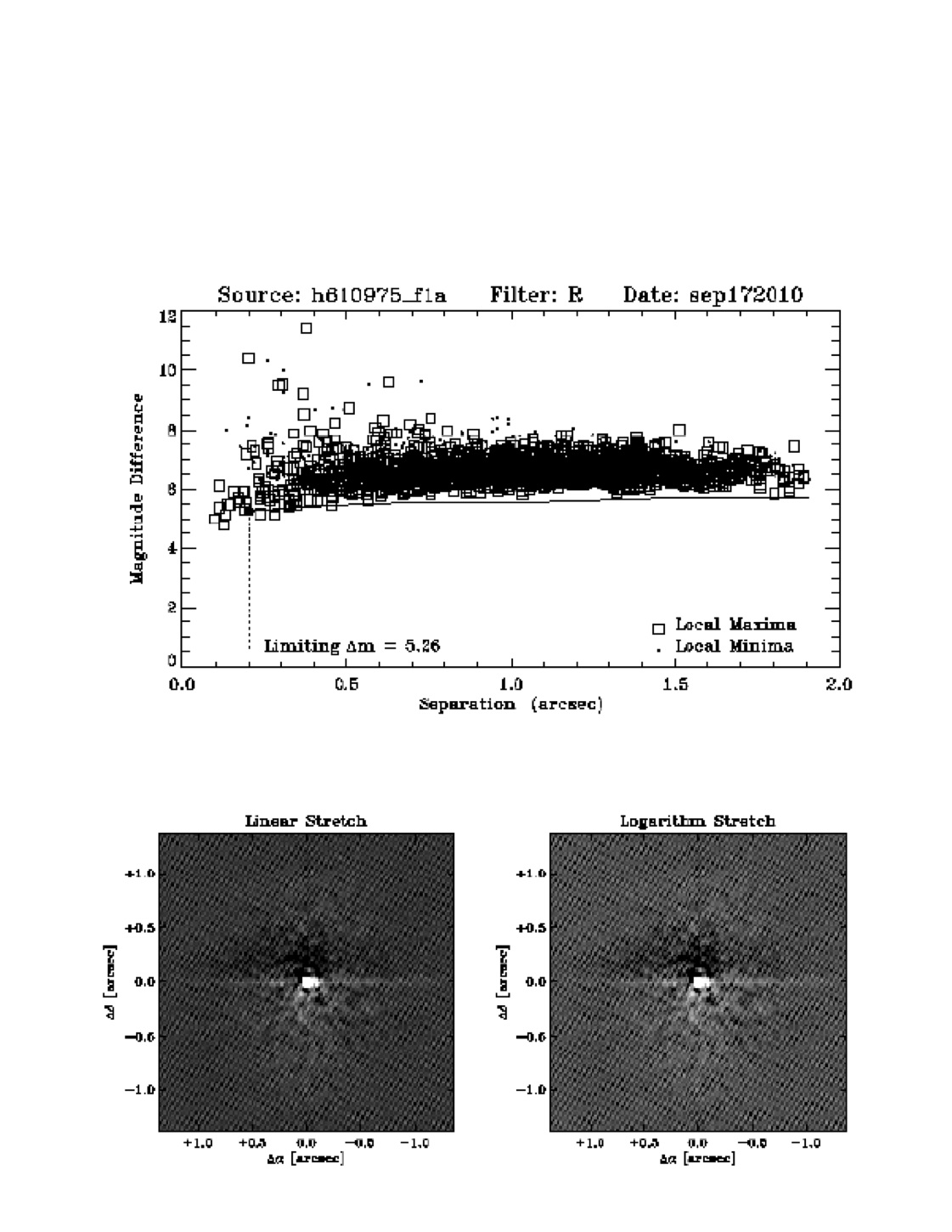}
\caption{Speckle observation of HD 179070
showing that no line-of-sight or real companions exist
from 0.05 to 2.8 arcsec of the star to a limit of 5.3 magnitudes in R (5.0 magnitudes in V) 
fainter than the star itself.
The reconstructed images at the bottom of the plot have N up and E left. The
horizontal line in the top plot shows the 5 sigma detection limit for companions
against the sky background (open squares) 
and the vertical line at 0.2 arc seconds is added to show the inner
limit for conservative multi-fringe speckle detections. 
}
\end{figure}


\begin{figure}
\includegraphics[angle=0,scale=0.6,keepaspectratio=true]{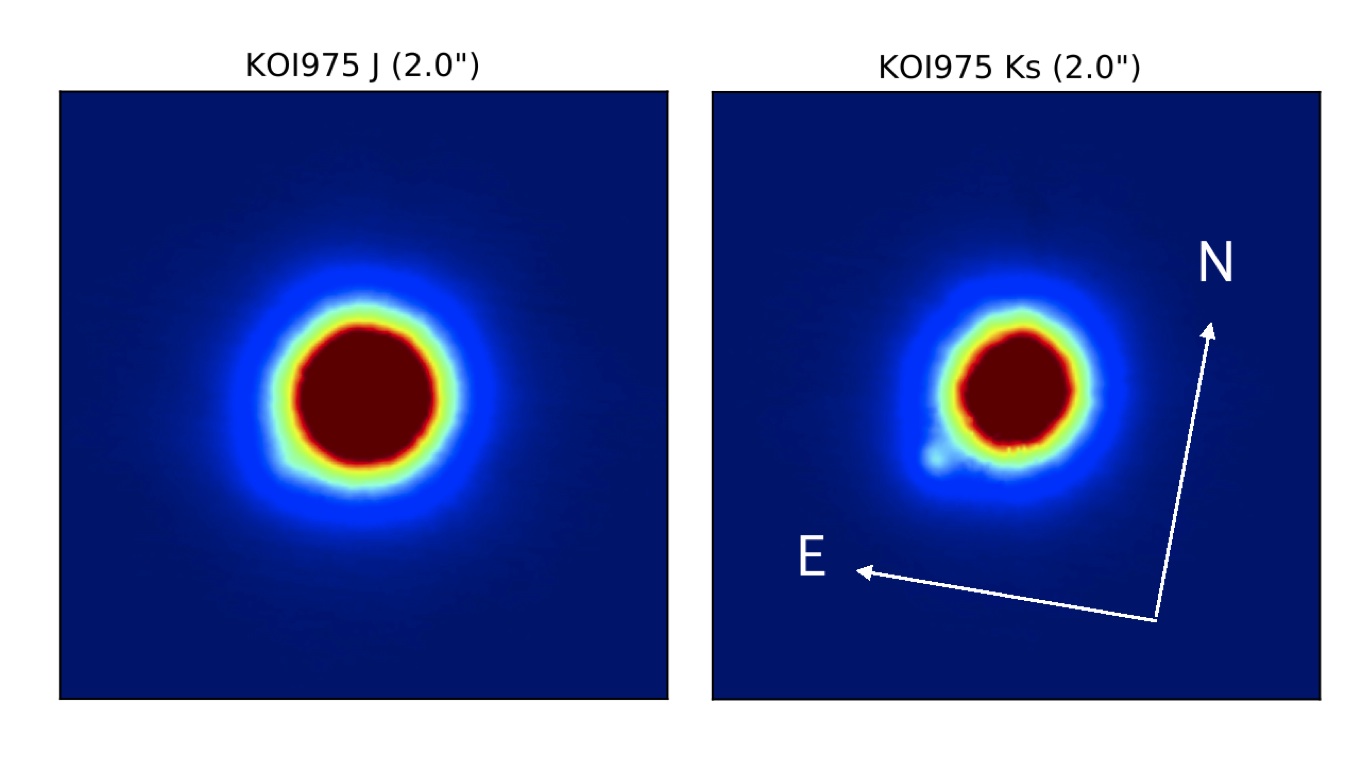}
\includegraphics[angle=90,scale=0.7,keepaspectratio=true]{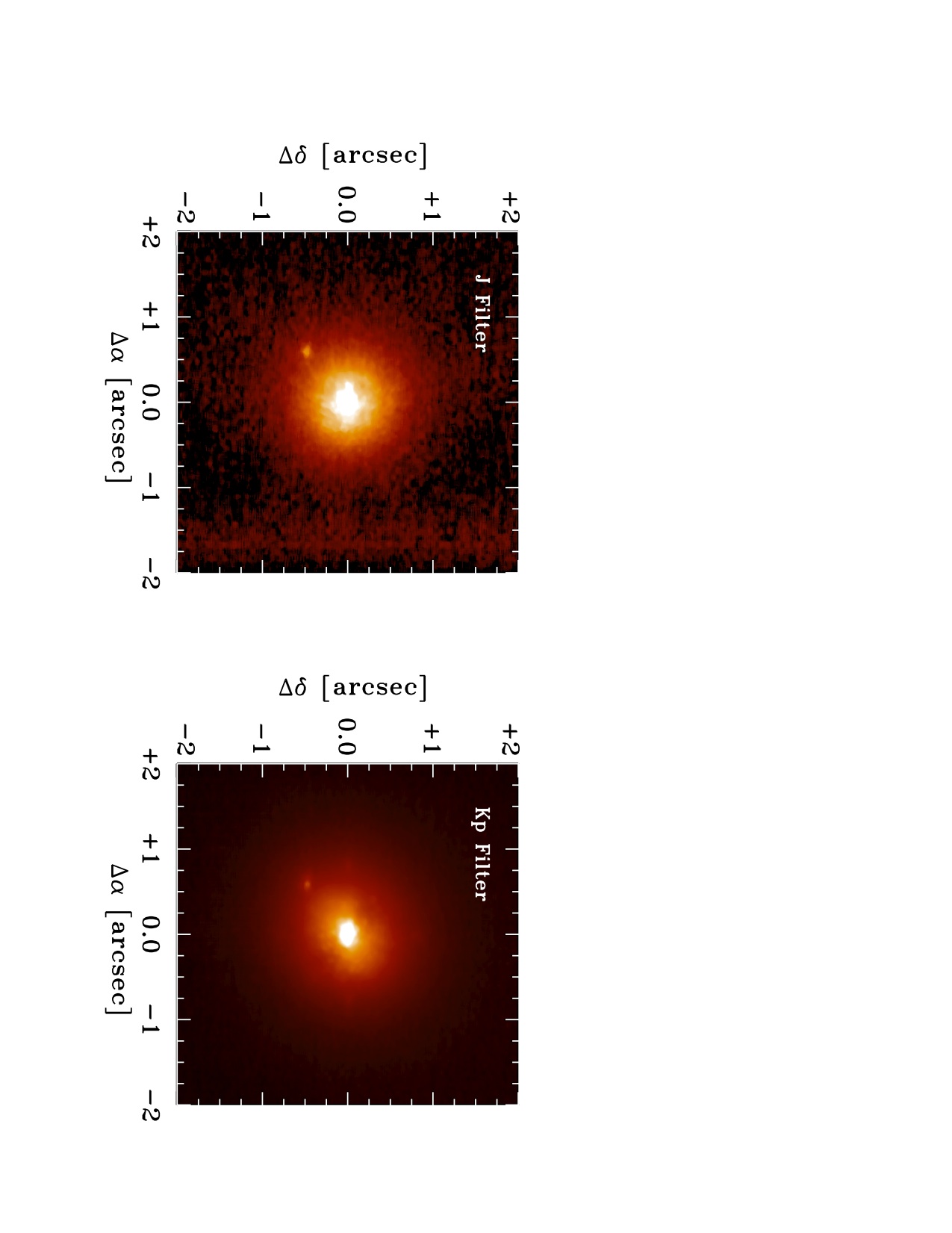}
\figcaption{
(top) ARIES AO images of HD 179070 in $J$ (left) and $Ks$ (right).
The inner 2" are shown, along with approximate N-E axes. The companion
star is clearly visible to the south-east of the main star in $Ks$,
and is suggested by a slight bump at the same location in $J$. No
other stars are seen within 10".
(bottom) $J$ and $K^\prime$ Keck-NIRC2 adaptive optics images of Kepler-21b.  The images
are centered on the primary target; the faint companion 
star can be seen approximately $0.75\arcsec$ to the SE of the target. 
\label{fig:keckAO} }
\end{figure}

%

\begin{figure}
\includegraphics[angle=0,scale=0.6,keepaspectratio=true]{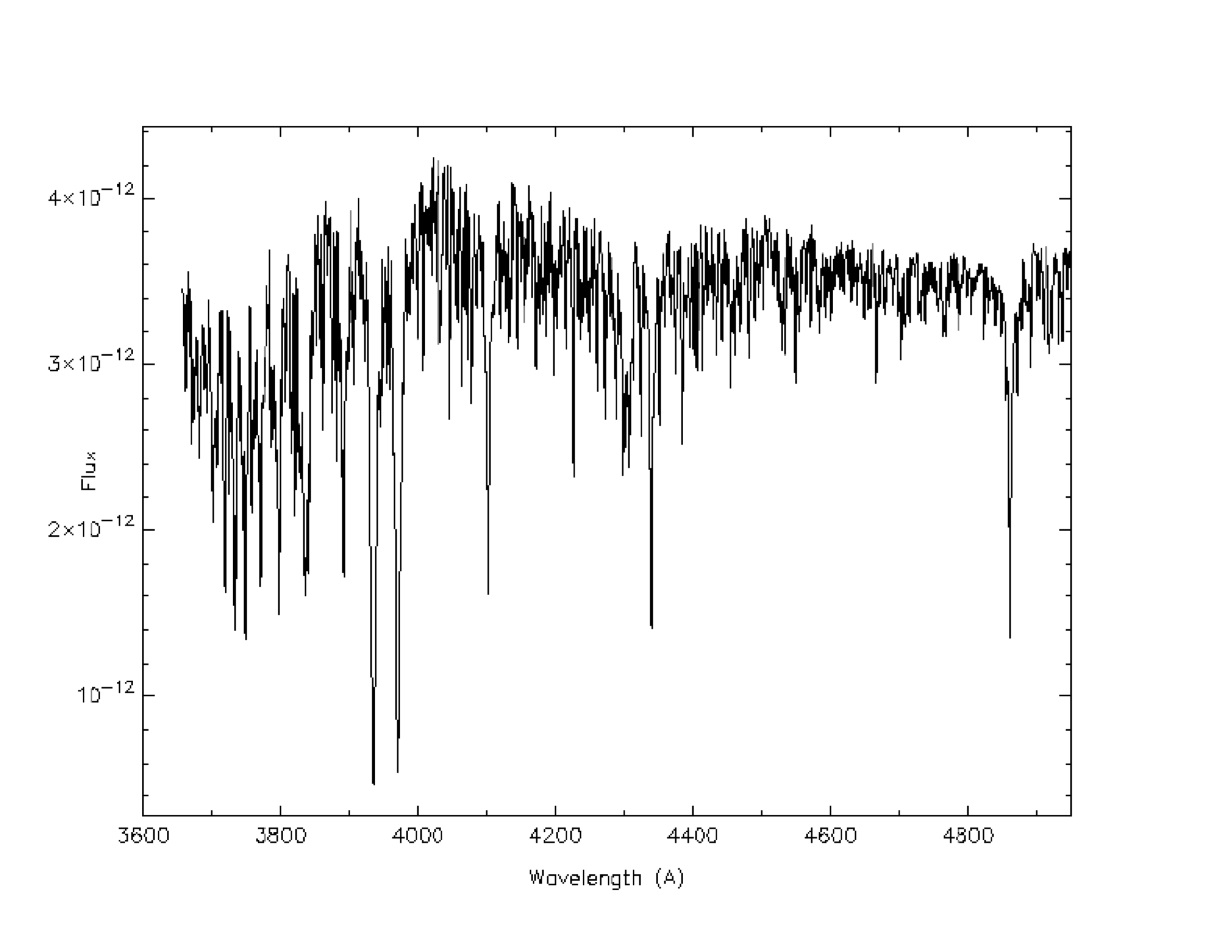}
\caption{The Kitt Peak 4-meter spectrum of HD 179070 obtained on 16 September 2010 UT.
The F6 IV star has an effective temperature of 6131K and a log g of 4.0. See Table 2.
}
\end{figure}

\begin{figure}
\includegraphics[angle=0,scale=0.6,keepaspectratio=true]{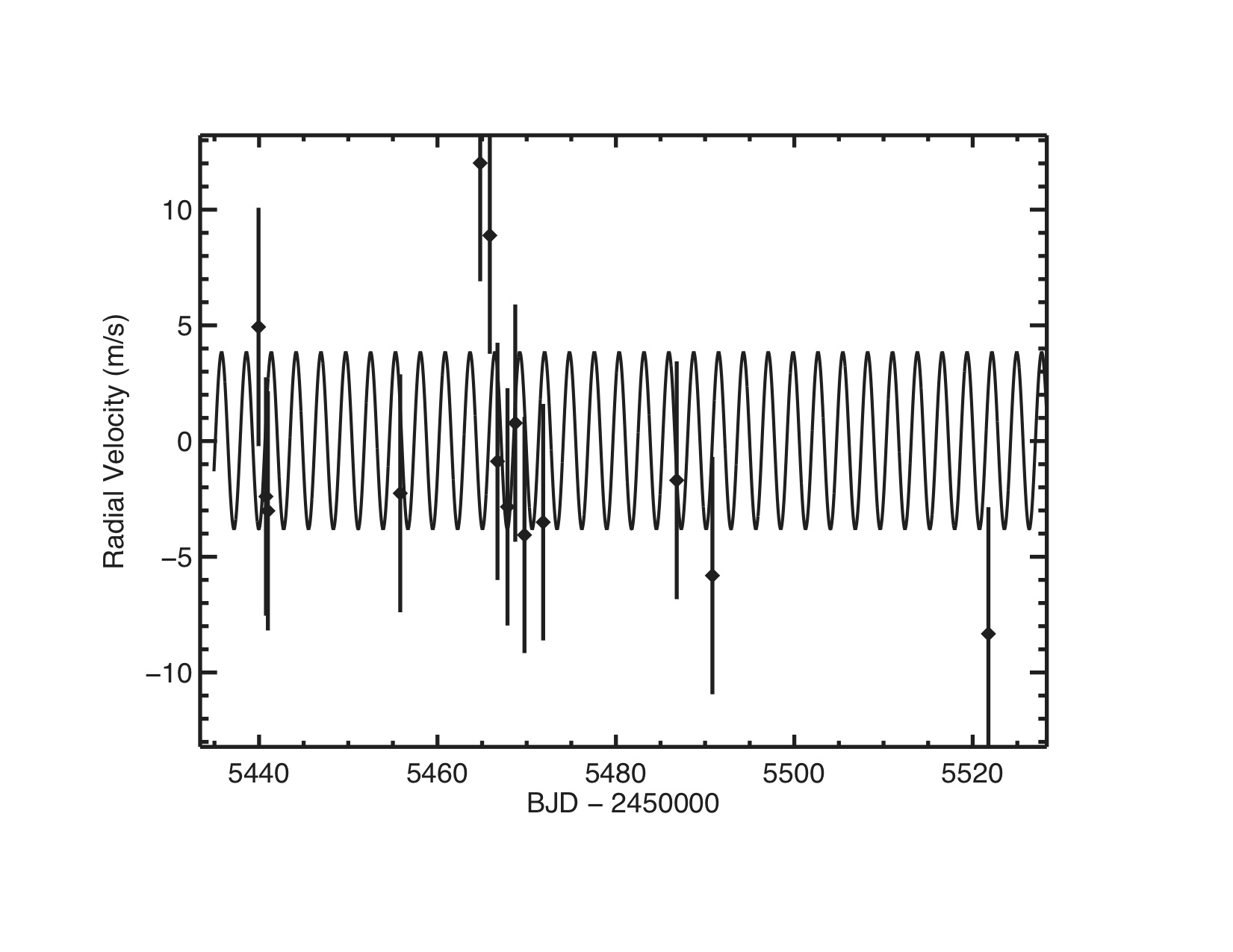}
\caption{
Radial velocity measurements for HD179070 from the Keck HIRES spectrometer are presented as a function of time. 
Internal errors of $\sim$2m/s are
added in quadrature to 5m/s of jitter to account for uncertainty in the measurements due to spectral type. 
A Keplerian orbit for Kepler-21b is overplotted (see test). The 
radial velocity amplitude does not correlate to the expected phase of the planet nor is any additional 
coherent variation observed.
The  radial velocities therefore provide only an upper limit to the amplitude. The small RMS
scatter of 5 m s$^{-1}$ imposes an upper limit on the planet mass (see text) and rules out a grazing incidence eclipsing binary.
}
\end{figure}

\begin{figure}
\includegraphics[angle=0,scale=0.6,keepaspectratio=true]{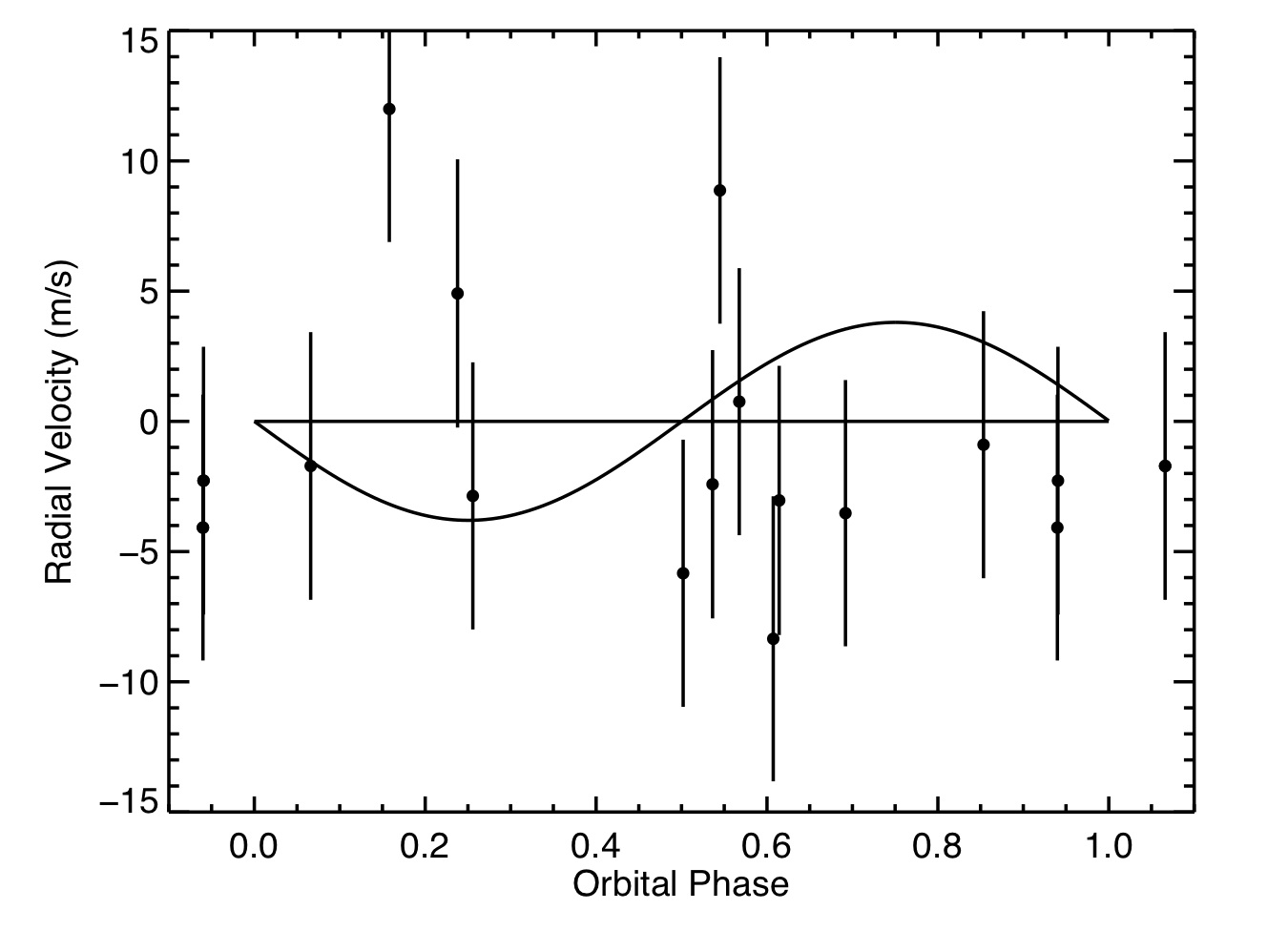}
\caption{
Keck velocities vs orbital phase for HD 179070. The velocities exhibit no evidence of coherence 
with orbital
phase, thus offering only a 3$\sigma$ upper limit to the mass of the planet of $\sim$20 Earth-masses.
The solid line shows the expected RV curve for a 10.4 Earth-mass planet orbiting HD 179070 (see \S8).
}
\end{figure}

\begin{figure*}
\epsscale{1.0}
\includegraphics[angle=0,scale=0.6,keepaspectratio=true]{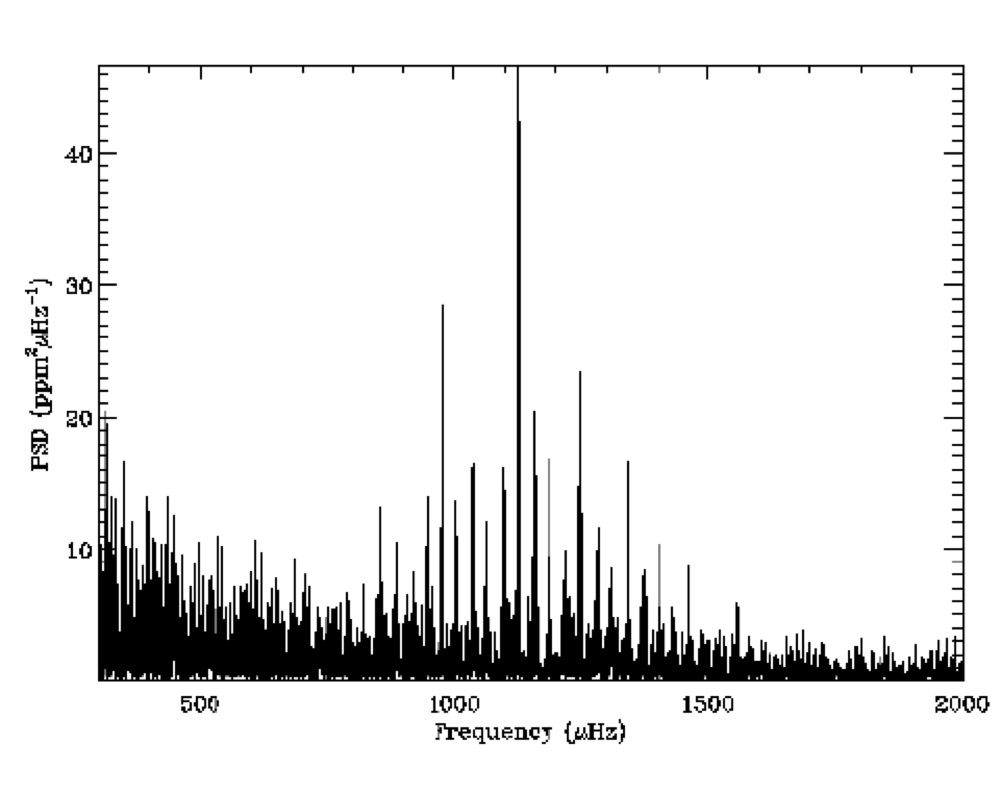}
\caption{Frequency-power spectrum of HD 179070, showing a rich pattern
 of overtones of solar-like oscillations. The rising background
 toward lower frequencies is due to convective granulation.}
\label{fig:powspec}
\end{figure*}

\begin{figure}
\includegraphics[angle=0,scale=0.6,keepaspectratio=true]{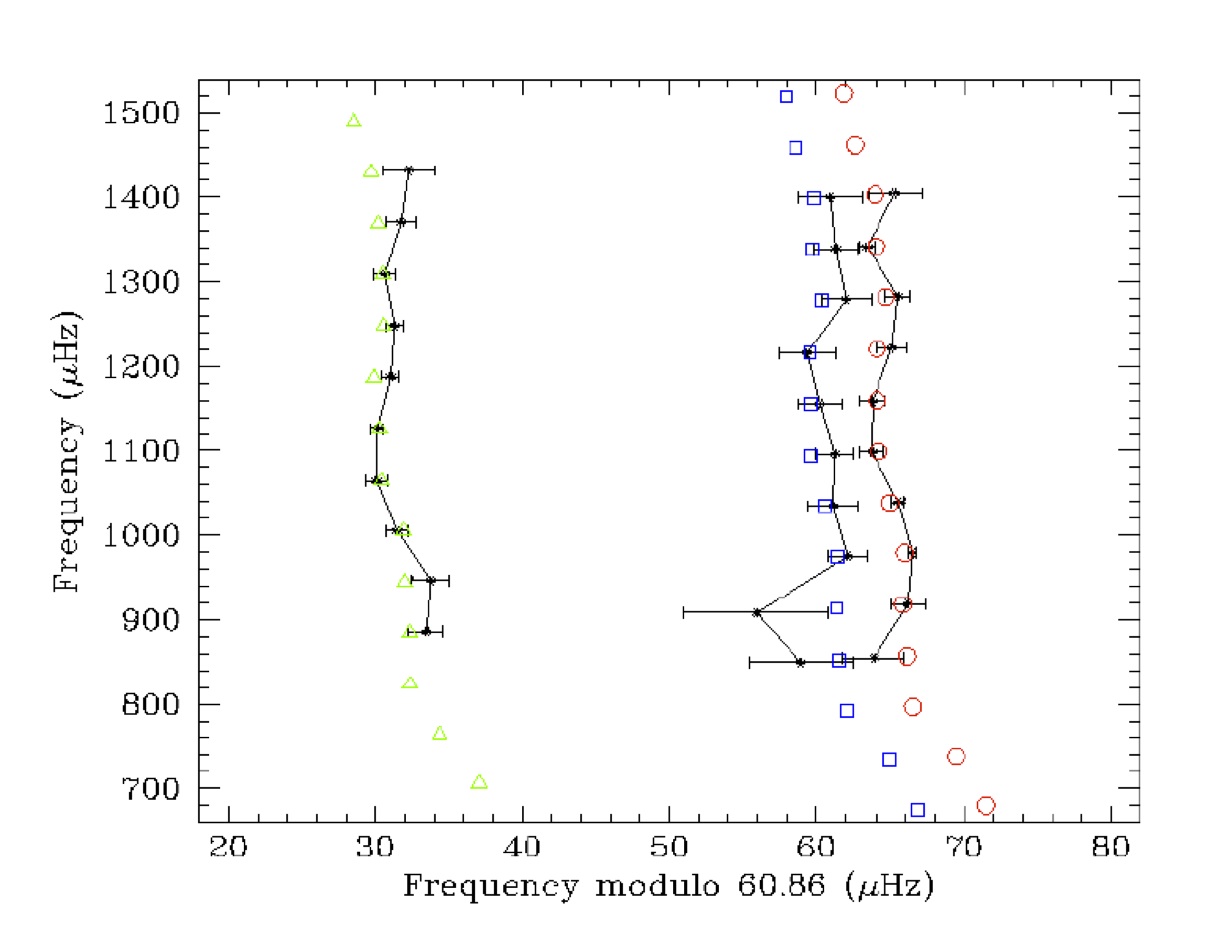}
\caption{Echelle diagram showing the observed frequencies from
Table~\ref{tab:freqs}, along with their $1\sigma$ uncertainties (black
symbols with error bars), and the best-fitting model frequencies
(color). Different symbol styles denote different spherical degrees,
$l$, circles showing $l=0$, triangles $l=1$ and squares $l=2$.}
\label{fig:ech}
\end{figure}


\begin{figure}
\includegraphics[angle=0,scale=0.7,keepaspectratio=true]{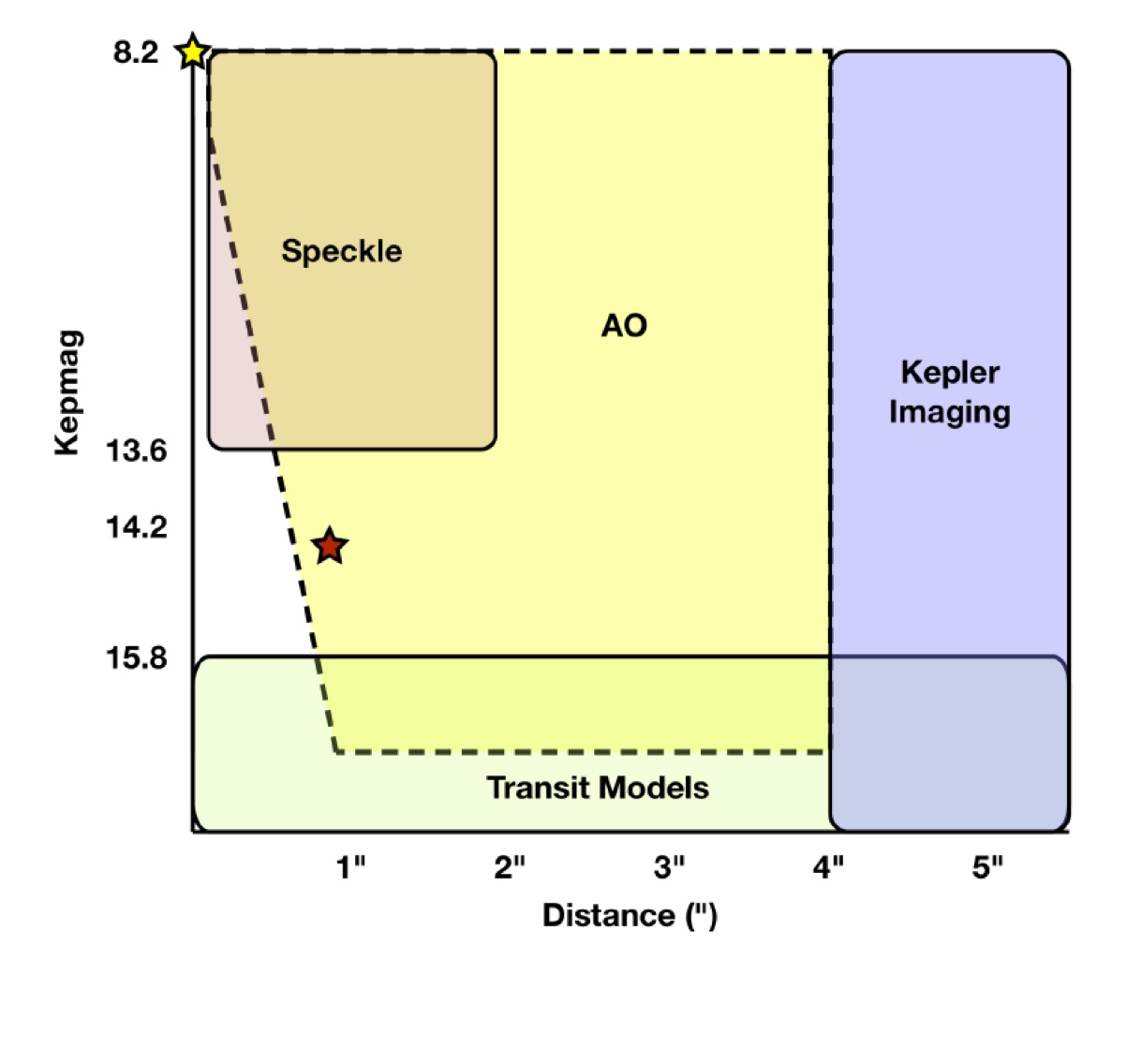}
\figcaption{
A schematic that demonstrates various techniques that rule 
out the brightness of potential blends in the Kepler aperture. The 
yellow star in the upper left corner represents HD 179070. The red star 
represents Kepmag=14.5 the companion discovered at a separation of 0.7" with 
AO.  This star was just undetected with speckle imaging due to the late 
stellar type.  Modeling of the transiting object rules out a blend by any star 
fainter than Kepmag=15.8 and a conservative estimate of the Kepler 
centroids eliminates any blend with a separation greater than 4".
\label{fig:koi975-blends} }
\end{figure}

\begin{figure}
\vskip 10pt
\begin{center}
\epsscale{0.8}
\includegraphics[angle=0,scale=0.7,keepaspectratio=true]{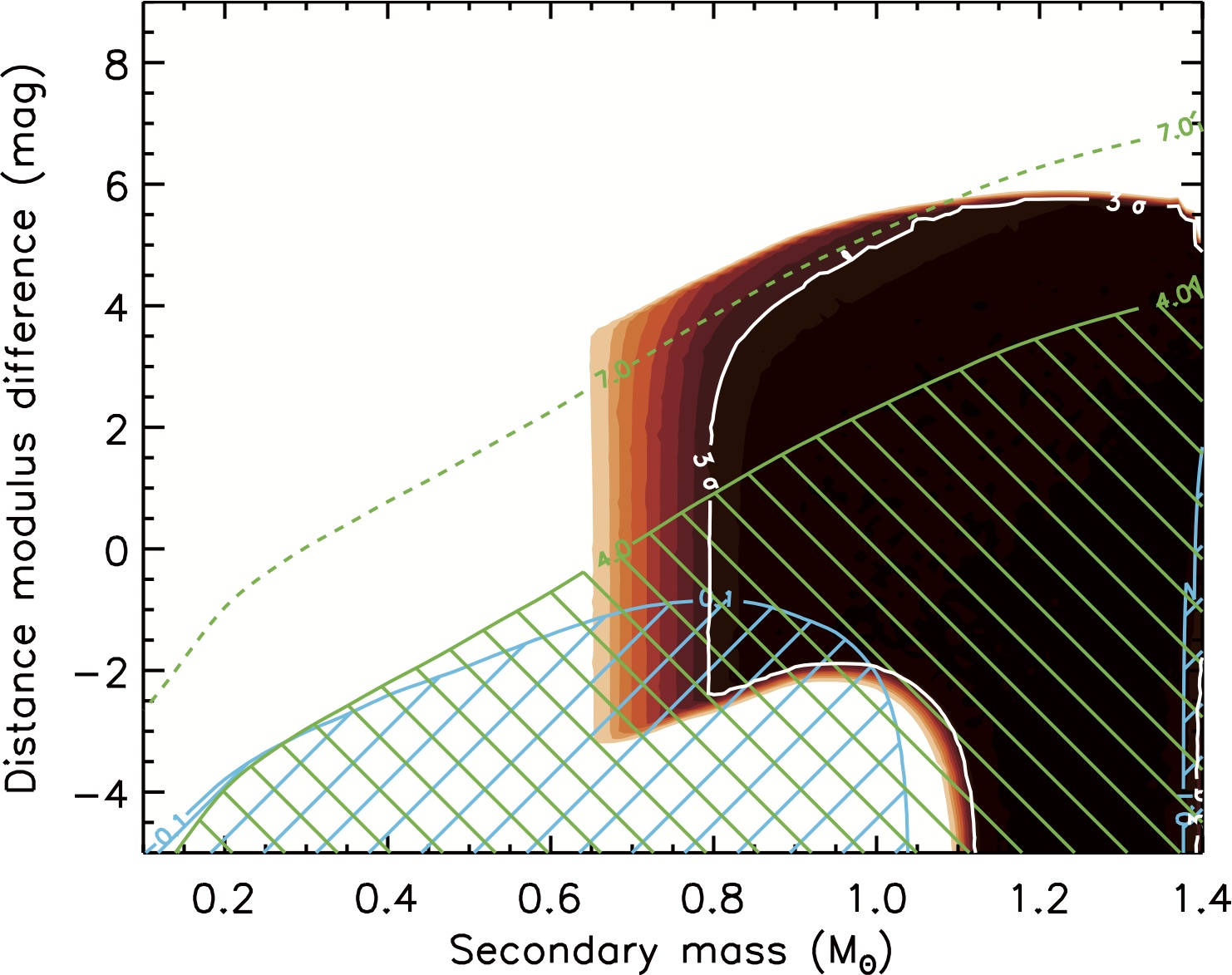}
\caption{Map of the $\chi^2$ surface (goodness of fit) for blends
involving background stars transited by a larger planet. The vertical
axis represents the distance between the background star and the
target \hd, expressed in terms of the difference in the distance
modulus.  Only blends inside the solid white contour match the
\kepler\ light curve within acceptable limits (3$\sigma$, where
sigma is the significance level of the $\chi^2$ difference compared to
a transit model fit; see Fressin et al. 2011).  Lighter-colored areas (red,
orange, yellow) mark regions of parameter space giving increasingly
worse fits to the data (4$\sigma$, 5$\sigma$, etc.), and correspond to
blends we consider to be ruled out.  The hatched blue region on the
lower left corresponds to blends that can be excluded because of their
overall $r\!-\!K_s$ colors, which are too red compared to the measured
index for \hd, by more than 3$\sigma$ (0.10~mag). A smaller similar
region is visible on the right. Blends that are bright enough to have
been detected spectroscopically are indicated by the hatched green
area, corresponding to contaminating stars that are up to 3~mag
fainter than the target. The faintest blends that remain can be as
much as 7 mag dimmer than the target (dashed green line).
\label{fig:blender_bp}}
\end{center}
\end{figure}

\begin{figure}
\vskip 10pt
\begin{center}
\epsscale{0.8}
\includegraphics[angle=0,scale=0.7,keepaspectratio=true]{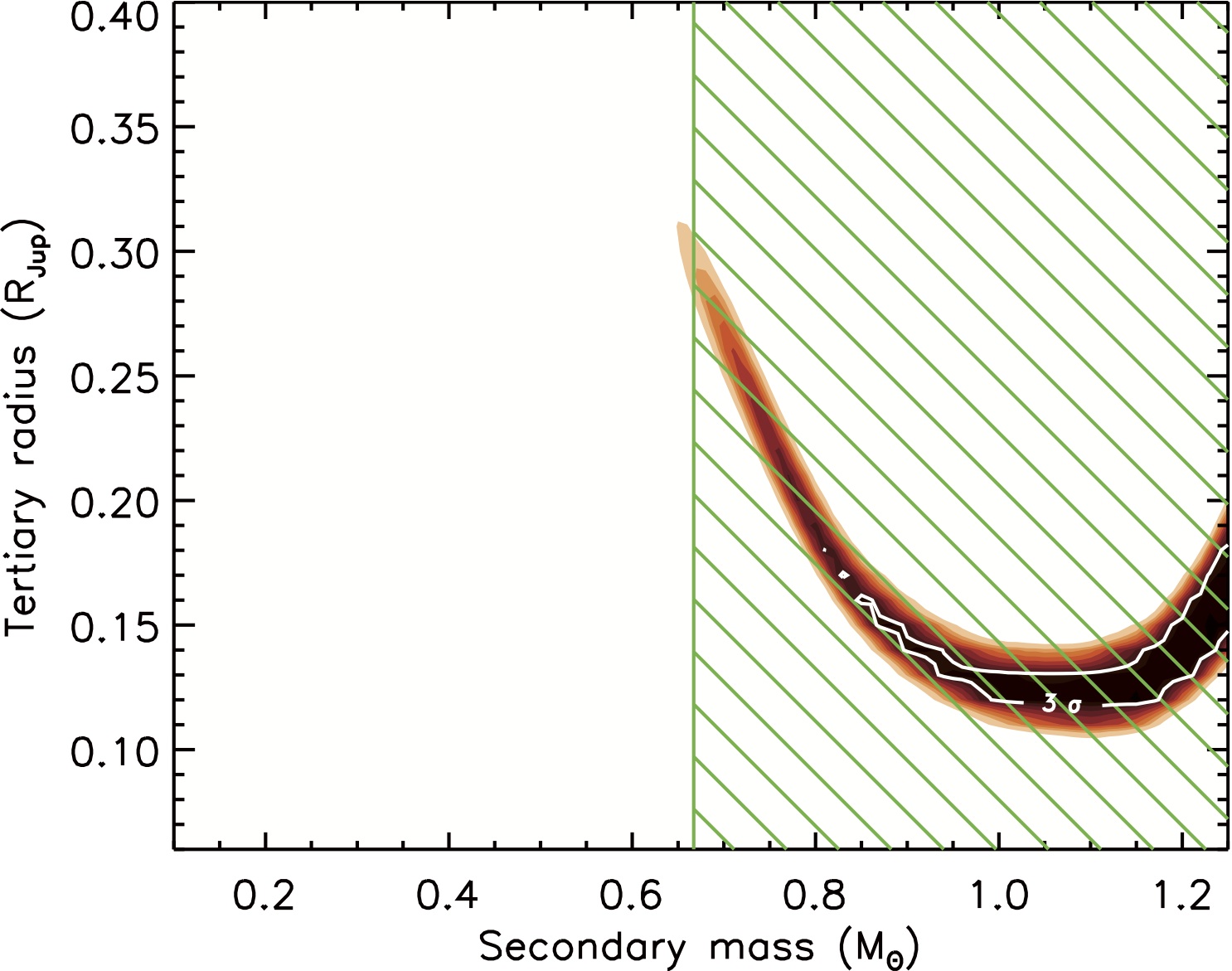}
\caption{Similar to Figure~\ref{fig:blender_bp} for the case of
hierarchical triple systems in which the secondary star is transited
by a planet, mimicking the signal in \hd.  In this case the vertical
axis represents the radius of those planets. Blends inside the white
3-$\sigma$ contour have light curves that match the shape observed for
\hd. While the $r\!-\!K_s$ colors of these blends are
indistinguishable from that of a single star like \hd\ at the
3-$\sigma$ level, the stars involved are all bright enough that most
would have been detected in our high-resolution spectra as a second
set of lines. This is indicated by the green hatched area. Only those
with a RV such that the lines are completely blended with those of the
target would escape notice (see text).
\label{fig:blender_htp}}
\end{center}
\end{figure}


\begin{figure}
\includegraphics[angle=0,scale=0.7,keepaspectratio=true]{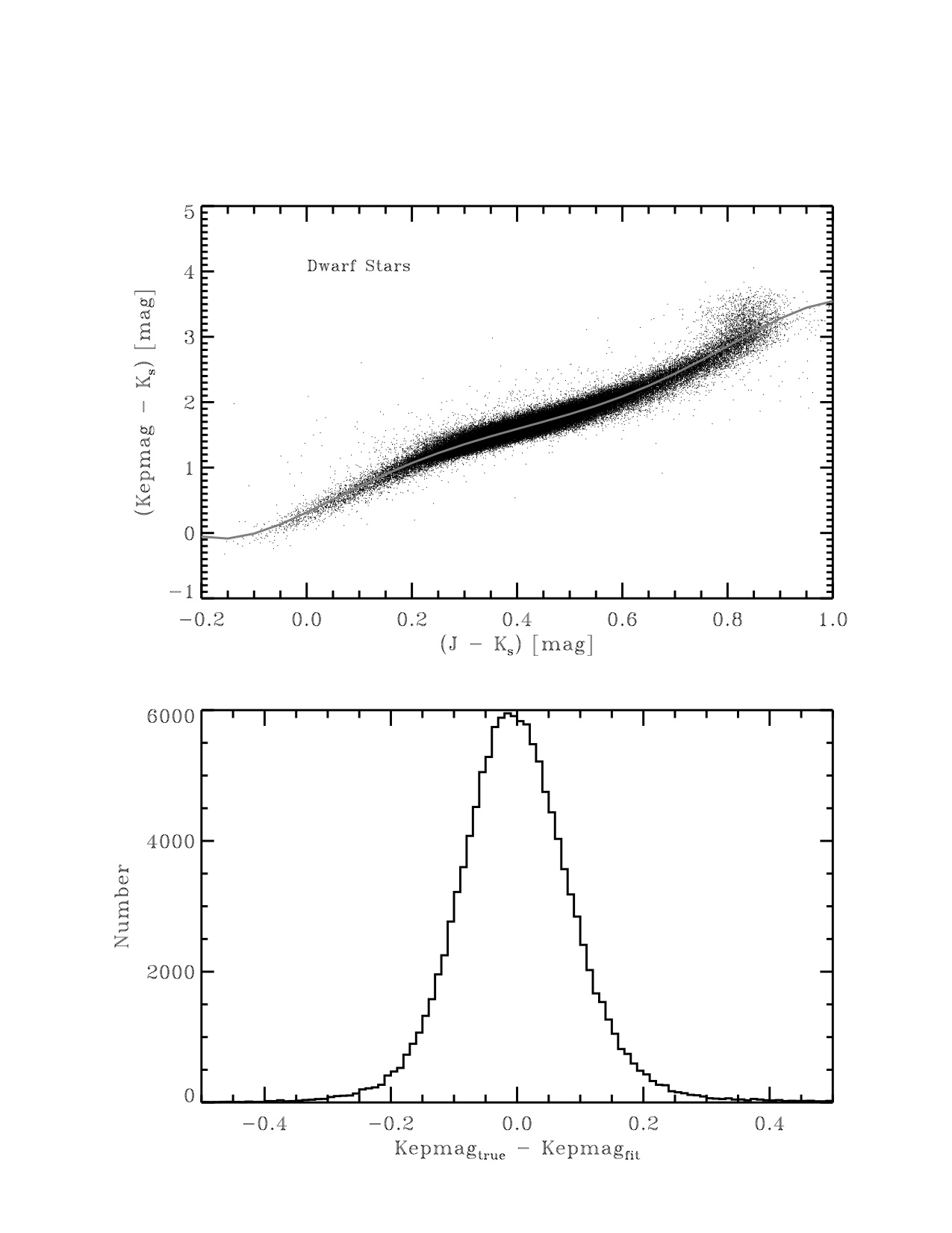}
\figcaption{
Top: $Kepmag-Ks$ vs $J-Ks$ for Q1 dwarf stars.  The grey line represents
the 5th order polynomial fit.  Bottom: histogram of the residuals from 
the polynomial fit.  
\label{fig:ccplot-dwarf} }
\end{figure}

\begin{figure}
\includegraphics[angle=0,scale=0.7,keepaspectratio=true]{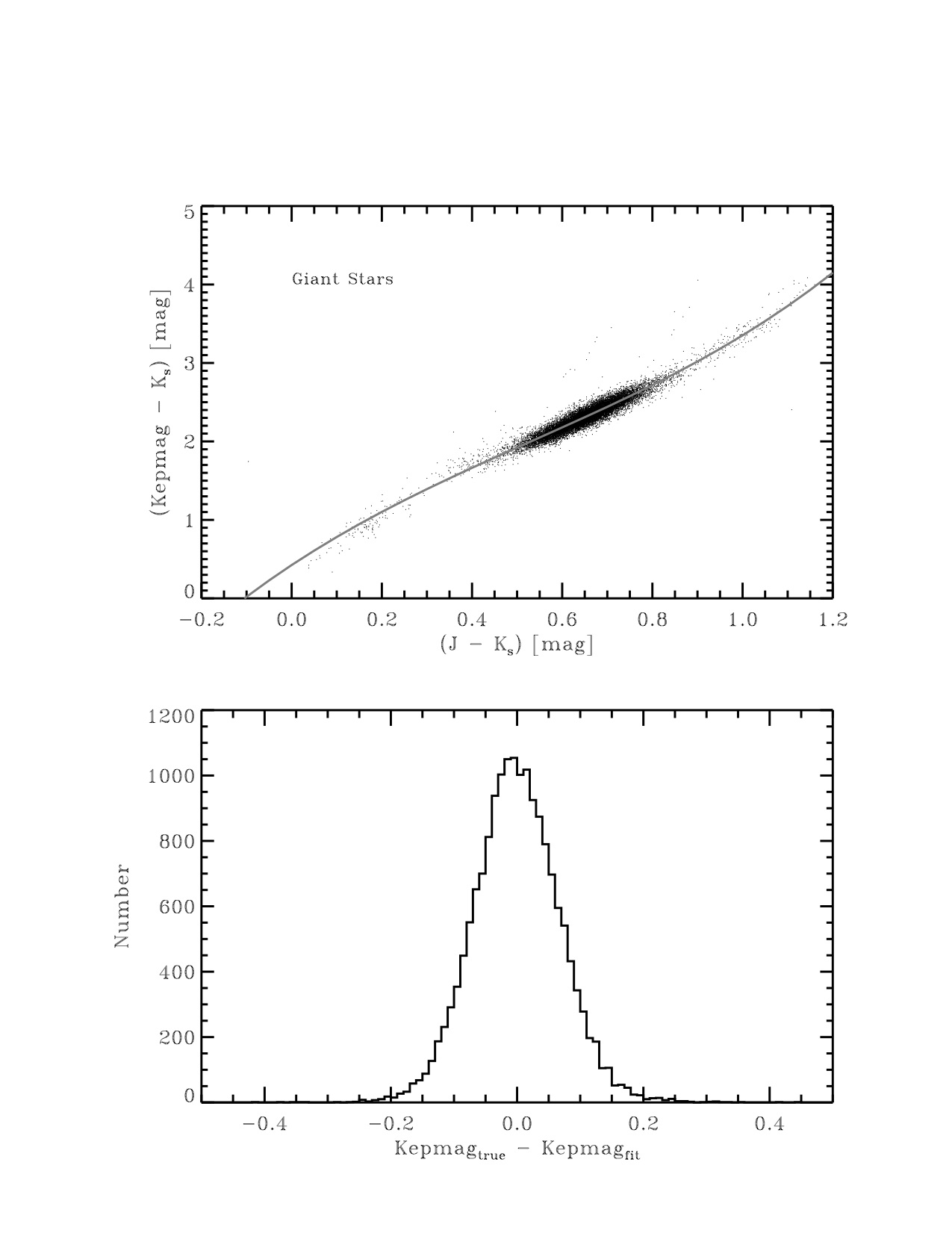}
\figcaption{
Top: $Kepmag-Ks$ vs $J-Ks$ for Q1 giant stars.  The grey line represents
the 3rd order polynomial fit.  Bottom: histogram of the residuals from 
the polynomial fit. 
\label{fig:ccplot-giant} }
\end{figure}

\begin{figure}
\includegraphics[angle=90,scale=0.5,keepaspectratio=true]{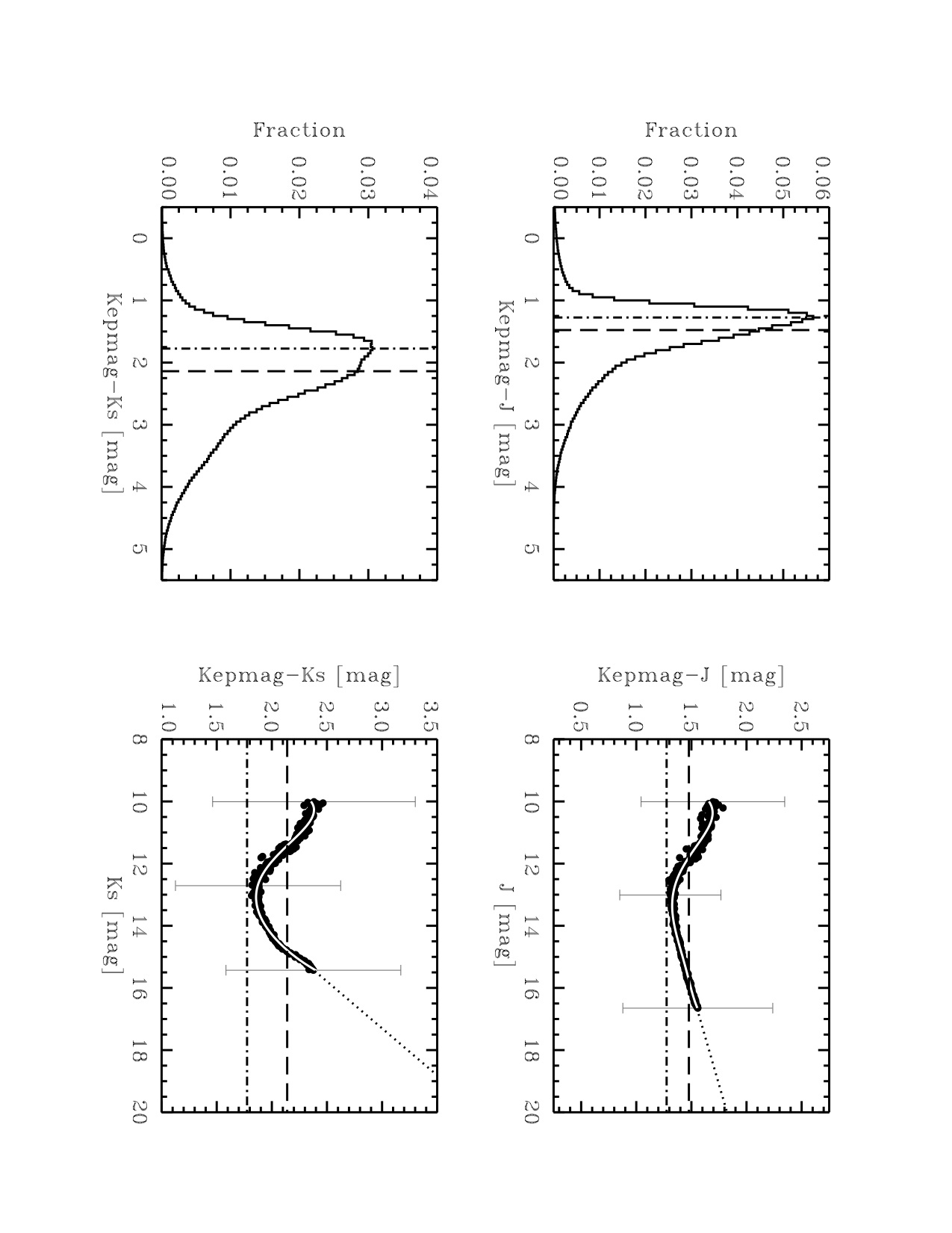}
\figcaption{ Top: (left) Histogram of the $Kp - J$ color for all
sources in the KIC.   The dashed line marks the median value and the
dashed-dot line marks the peak of the histogram.  (right) Median $Kp -
J$ color for all sources in the KIC as a function of $J$ mag (binsize = 0.02
mag in $J$).  The horizontal dashed and dashed-dot lines mark the same values
as  in the histogram. Three representative dispersions per bin are shown as
error bars.  The white solid line represents a 5th-order polynomial fit to
the data, and the dotted line represents a linear fit to sources fainter
than $J=15$ mag extrapolated to magnitudes beyond the limits of the KIC.
Bottom: (left) Histogram of the $Kp - Ks$ color for all
sources in the KIC. (right) Median $Kp -
Ks$ color for all sources in the KIC as a function of $Ks$ mag (binsize = 0.02
mag in $Ks$).  The various lines demark the same values as in the top plots but
for $Kp - Ks$ values.  \label{fig:kepmag_color} }
\end{figure}

\newpage

\begin{thebibliography}

\bibitem[x]{y} Appourchaux, T., Michel, E., Auvergne, M., et al., 2008,
A\&A, 488, 705

\bibitem[Ballard et al.(2011)]{Ballard:11}
 Ballard, S.\ et al.\ 2011, \apj, in press (arXiv:1109.1561)

\bibitem[x]{y} Basu, S., Chaplin, W. J., Elsworth, Y., 2010, ApJ, 710,
1596

\bibitem[Batalha et al.(2010)]{batalha10} Batalha, N., et al., 2010, \apj, 713,
L103

\bibitem[Batalha et al.(2010)]{batalha10} Batalha, N., et al., 2011, \apj, 729, 27

\bibitem[Borucki et al.(2010a)]{bor10a} Borucki, W., et al., 2010a, Science, 327, 977

\bibitem[Borucki et al.(2010b)]{bor10b} Borucki, W., et al., 2010b, \apj, 713, L126

\bibitem[Borucki et al.(2010b)]{bor10b} Borucki, W., et al., 2011, \apj, 736, 19

\bibitem[x]{y} Brown, T. M., Gilliland, R. L., Noyes, R. W., Ramsey,
 L. W., 1991, ApJ, 368, 599

\bibitem[Brown et al.(2011)]{Brown:11}
 Brown, T.\ M., Latham, D.\ W., Everett, M.\ E., \& Esquerdo, G.\
 A. 2011, \aj, 142, 112

\bibitem[Bryson et al.(2010)]{bry10} Bryson, S. T., et al., 2010, \apj,
713, L97

\bibitem[x]{y} Buchhave, L. A., et al. 2010, ApJ, 720, 1118

\bibitem[x]{y} Campante, T. L., Karoff, C., Chaplin, W. J., Elsworth,
 Y., Handberg, R., Hekker, S., 2010a, MNRAS, 408, 542

\bibitem[x]{y} Campante, T. L., Handberg, R., Mathur, S., et al. 2011,
A\&A, in press
\bibitem[x]{y} Chaplin, W. J., Appourchaux, T., Elsworth, Y., et al.,
 2010, ApJ, 713, L169

\bibitem[x]{y} Chaplin, W. J., Kjeldsen, H., \&  Christensen-Dalsgaard, J.,
 2011, Science, 332, 213

\bibitem[x]{y} Christensen-Dalsgaard, J.\ 1993, ASP Conf., 42, 347

\bibitem[x]{y} Christensen-Dalsgaard, J.\ 2008a, AP\&SS, 316, 13

\bibitem[x]{y} Christensen-Dalsgaard, J.\ 2008b, AP\&SS, 316, 113

\bibitem[x]{y} Christensen-Dalsgaard, J., Kjeldsen, H., Brown, T. M., et
 al., 2010, ApJ, 713, L164

\bibitem[Ciardi et al.(2011)]{crd11}Ciardi, D. R. et al. 2011, \aj, 
141, 108

\bibitem[x]{y} Creevey, O.~L., Monteiro, M.~J.~P.~F.~G., Metcalfe,
T.~S., Brown, T.~M., Jim{\'e}nez-Reyes, S.~J., \& Belmonte, J.~A.\
2007, ApJ, 659, 616

\bibitem[Cochran et al.(2011)]{Cochran:11}
 Cochran, W.\ D.\ et al.\ 2011, \apjs, in press (arXiv:1110.0820)

\bibitem[dem(2005)]{for05} Demarque, P., et al., 2004, \apj, 155, 667

\bibitem[Ford(2005)]{for05} Dunham, E. W., et al., 2010, \apj, 713, L136

\bibitem[x]{y} F\H{u}r\'esz, G.~2008, Ph.D.~thesis, University of Szeged, Hungary

\bibitem[Fink et al.] {y} Finkbeiner, D.~P., \& Davis, M.\ 1998, \apj, 500, 525 

\bibitem[x]{y} Fletcher, S. T., Broomhall, A.-M., Chaplin, W. J., et
   al., 2010, MNRAS, in the press

\bibitem[Ford(2005)]{for05} Ford, E., 2005, \aj, 129, 1706

\bibitem[Fressin et al.(2011)]{Fressin:11}
 Fressin, F.\ et al.\ 2011, \apj, in press

\bibitem[x]{y} Gai, N., Basu, S., Chaplin, W. J., Elsworth, Y., 2011,
 ApJ, submitted (arXiv:1009.3018G)

\bibitem[x]{y} Garc\'ia, R. A., Hekker, S., Guti\'errez-Soto, J., et
al., 2011, MNRAS, 414, 6

\bibitem[x]{y} Gelman, A., \& Rubin, D. B. 1992, Stat. Sci., 7, 457

\bibitem[x]{y} Gilliland, R. L., et al., 2010, ApJ, 713, L160

\bibitem[x]{y} Gilliland, R. L., Brown, T. M., Christensen-Dalsgaard, J.,
 et al., PASP, 2010, 122, 131

\bibitem[x]{y} Gregory, P., 2011, MNRAS, 410, 94

\bibitem[x]{y} Gould, B. A. 1855, AJ, 4, 81

\bibitem[x]{y} Grec, G., Fossat, E., Pomerantz, M., 1983, SolPhys, 82,
 55

\bibitem[x]{y} Hekker, S., Broomhall, A.-M., Chaplin, W. J., et al.,
2010, MNRAS, 402, 2049

\bibitem[Horch et al.(2010)]{horch10} Holman, M. et al., 2010, Science, 330, 51

\bibitem[Horch et al.(2010)]{horch10} Horch, E. P. et al., 2010, \aj, 141, 45

\bibitem[Howell et al.(2010)]{how10} Howell, S. B., et al., 2010, \apj, 725, 1633

\bibitem[Howell et al.(2011)]{how11} Howell, S. B., et al., 2011, \aj, 142, 19

\bibitem[x]{y} Huber, D., Stello, D., Bedding, T. R., et al., 2009,
CoAst, 160, 74

\bibitem[x]{y} Isaacson, H. \& Fischer, D. A., 2010, ApJ, 725, 885

\bibitem[Jacoby et al.(1984)]{jac84} Jacoby, G. H., Hunter, D., and
Christian, C., 1984, ApJ Suppl., 56, 257

\bibitem[Jenkins et al.(2010a)]{jen10a} Jenkins, J. M., et al., 2010a, \apj, 724, 1108

\bibitem[Jenkins et al.(2010b)]{jen10b} Jenkins, J. M., et al., 2010b,
\apj, 713, L87 

\bibitem[Jenkins et al.(2010c)]{jen10c} Jenkins, J. M., et al., 2010c,
\apj, 713, L120 

\bibitem[x]{y} Jenkins, J., et al., 2011, \apj, submitted 

\bibitem[x]{y} Karoff, C., Campante, T. L., Chaplin, W. J., 2010, AN, 331, 972

\bibitem[x]{y} Kjeldsen, H., Bedding, T. R., 1995, A\&A, 293, 87

\bibitem[x]{y} Kjeldsen, H., et al.\ 2008, ApJL, 683, L175

\bibitem[x]{y} Koch, D. G., et al., 2010a, ApJ, 713, L131
\bibitem[x]{y} Koch, D. G., et al., 2010b, ApJ, 713, L79

\bibitem[x]{y} Latham, D. W., et al., 2010, \apj, 713, L140

\bibitem[Leggett et al.(2002)]{2002ApJ...564..452L} Leggett, S.~K., et al.\ 
2002, \apj, 564, 452 

\bibitem[x]{y} Lissauer, J. J., et al., 2011, Nature, 470, 53

\bibitem[Mandel \& Agol(2002)]{man02} Mandel, K. \& Agol, E., 2002, \apj, 580,
171

\bibitem[x]{y} Marcus, R., et al., 2010, ApJ, 712, L73.

\bibitem[x]{y} Marcy, G. W., et al., 2008, Physica Scripta Vol. T, 130, 014001

\bibitem[x]{y} Mathur, S., Garc\'\i a, R. A., R\'egulo C., et al., 2010a,
A\&A, 511, 46

\bibitem[x]{y} Mathur, S., Handberg, R., Campante, T. L., et al., 2011,
ApJ, 733, 95

\bibitem[x]{y} Metcalfe, T.~S., \& Charbonneau, P.\ 2003,
J.~Computat.~Phys., 185, 176

\bibitem[x]{y} Metcalfe, T.~S., et al.\ 2009, ApJ, 699, 373

\bibitem[x]{y} Metcalfe, T. S., Monteiro M. J. P. F. G., Thompson,
 M. J., et al., 2010, ApJ, 723, 1583

\bibitem[MZ]{y} Molenda-Zakowicz et al., 2011, MNRAS, 412, 1210

\bibitem[Morton \& Johnson(2011)]{Morton:11}
 Morton, T.\ D., \& Johnson, J.\ A. 2011, \apj, 738, 170

\bibitem[x]{y} Mosser, B., Appourchaux, T., 2009, A\&A, 508, 877

\bibitem[Nord(2004]{n04} Nordstr\"{o}m, B., et al. 2004, A\&A, 418, 989

\bibitem[x]{y} Peirce, B. 1852, AJ, 2, 161

\bibitem[x]{y} Piskunov, N. \ 1996, \aaps, 118, 595                                                          
\bibitem[x]{y} Quirion, P.-O., Christensen-Dalasgaard, J., Arentoft, T.,
 2010, ApJ, 725, 2176

\bibitem[Raghavan et al.(2010)]{Raghavan:10}
 Raghavan, D.\ et al.\ 2010, \apjs, 190, 1

\bibitem[Robin et al.(2003)]{Robin:03}
 Robin, A.\ C., Reyl\'e, C., Derri\'ere, S., \& Picaud, S. 2003, \aap,
 409, 523

\bibitem[Rowe et al.(2006)]{row06} Rowe, J. F., et al. 2006, \apj, 646, 1241


\bibitem[x]{y} Roxburgh, I. W., 2009, A\&A,
506, 435

\bibitem[Torres et al.(2004)]{Torres:04}
 Torres, G., Konacki, M., Sasselov, D.\ D., \& Jha, S.\ 2004, \apj,
 614, 979

\bibitem[Torres et al.(2010)]{torres10} Torres, G., et al., 2011, \apj, 727, 24

\bibitem[Schlegel et al.(1998)]{1998ApJ...500..525S} Schlegel, D.~J., et al., 1998, \apjs, 500, 525

\bibitem[Schuler et al.(2011)]{2011ApJ...732...55S} Schuler, 
S.~C., Flateau, D., Cunha, K., et al.\ 2011, \apj, 732, 55

\bibitem[x]{y} Stello, D., Chaplin, W. J., Bruntt, H., 2009, MNRAS, 700,
1589

\bibitem[Tokunaga \& Vacca(2005)]{tv05}Tokunaga, A. T. \& Vacca, W. D.
2005, \pasp, 117, 421

\bibitem[x]{y} Valencia, D., et al., 2007, ApJ, 665, 1413

\bibitem[Valenti \& Fischer(2005)]{Valenti05} Valenti, J.~A.,\& Fischer, D.~A.\ 2005, \apjs, 159, 141

\bibitem[Van]{y} van Cleve, J., 2008, {\it Kepler} Instrument Handbook (available at
http://archive.stsci.edu/kepler/)

\bibitem[Van]{y} van Cleve, J., (ed.) 2009, {\it Kepler} Data release Notes (available at
http://archive.stsci.edu/kepler/)

\bibitem[x]{y} Verner, G. A., Elsworth, Y., Chaplin, W. J., et al.,
 2011, MNRAS, in press

\bibitem[x]{y} Vogt, S. S., et al., 1994, Proc. SPIE, 2198, 362

\bibitem[x]{y} Welsh, W., et al., 2011, \apj, arXiv1102.1730

\bibitem[x]{y} White, T. R., Bedding, T. R., Stello, D., 2011, ApJ, submitted

\bibitem[x]{y} Woitaszek, M., Metcalfe, T. S., Shorrock I., 2009, in:
 Proceedings of the 5th Grid Computing Environments Workshop (New
 York: ACM), 1 - 7, doi:10.1145/1658260.1658262

\end{thebibliography}
\end{document}